\documentclass[12pt]{article}

\usepackage{amsmath}
\usepackage{amssymb}
\usepackage{amsfonts}
\usepackage{eucal}
\usepackage{dsfont}
\usepackage{epsfig}
\usepackage{graphicx}
\usepackage{mathrsfs}
\usepackage{dsfont}

\newcommand{\be}{\begin{equation}}
\newcommand{\bea}{\begin{eqnarray}}
\newcommand{\eea}{\end{eqnarray}}
\newcommand{\ba}{\begin{array}}
\newcommand{\ea}{\end{array}}
\newcommand{\ee}{\end{equation}}

\expandafter\ifx\csname mathbbm\endcsname\relax

\else

\fi

\newcommand{\Tr}{{{Tr}}}

\newcommand{\supq}{{U(N)^{p \times q}}}

\newcommand{\neqone}{{\mathcal{N}=1}}
\makeatletter \@addtoreset{equation}{section}

\makeatother

\def\l{\label}

\def\cN{{\cal N}}
\def\cD{{\mathfrak{D}}}
\def\cO{{\cal O}}
\def\orb{$AdS_5\times S^5/Z_p\times Z_q$}

\def\opt{operator}
\begin{document}

\begin{titlepage}
\hfill
\vbox{
    \halign{#\hfil         \cr
           SU-ITP-05/30\cr
           IPM/P-2005/037  \cr
           hep-th/0510189\cr
           } 
      }  
\vspace*{6mm}

\begin{center}

{\Large {\bf Integrable Spin Chains on the Conformal Moose\\
}} \vspace*{2mm}

{\large\bf{$\mathcal{N}=1$ Superconformal Gauge Theories as\\
Six-Dimensional String Theories}}

\vspace*{10mm} \vspace*{1mm}

{\bf Darius Sadri$^{1}$, M. M. Sheikh-Jabbari$^{2}$}

\vspace*{0.4cm}
{\it ${}^1${Department of Physics, Stanford University\\
382 via Pueblo Mall, Stanford CA 94305-4060, USA}}

\vspace*{0.4cm}

{\it ${}^2${Institute for Studies in Theoretical Physics and Mathematics (IPM)\\
P.O.Box 19395-5531, Tehran, Iran}}

\vspace*{0.2cm}

\vspace*{0.4cm}

{\tt darius@itp.stanford.edu,$\,$jabbari@theory.ipm.ac.ir}

\vspace*{1cm}
\end{center}

\begin{center}
{\bf\large Abstract}
\end{center}


We consider $\cN=1, \ D=4$ superconformal $U(N)^{p \times q}$
Yang-Mills theories dual to \orb\ orbifolds. We construct the
dilatation operator of this superconformal gauge theory at
one-loop planar level. We demonstrate that a specific sector of
this dilatation operator can be thought of as the transfer matrix
for a two-dimensional statistical mechanical system, related to
an integrable $SU(3)$ anti-ferromagnetic spin chain system, which in
turn is equivalent to a $2+1$-dimensional string theory where the
spatial slices are discretized on a triangular lattice. This is an
extension  of the $SO(6)$ spin chain picture of $\cN=4$ super
Yang-Mills theory. We comment on the integrability of this $\cN=1$
gauge theory and hence the corresponding three-dimensional
statistical mechanical system, its connection to three-dimensional
lattice gauge theories, extensions to six-dimensional string
theories, AdS/CFT type dualities and finally their construction
via orbifolds and brane-box models. In the process we discover a
new class of almost-BPS BMN type operators with large engineering
dimensions but controllably small anomalous corrections.


\end{titlepage}

\tableofcontents

\section{Introduction}

According to AdS/CFT \cite{AdS/CFT, Aharony:1999ti}
string theory on a given background (which is a gravitating
theory) is dual to a (usually non-gravitational) gauge theory. The
best known example of this duality is between type IIB strings on
$AdS_5\times S^5$ with $N$ units of the fiveform flux on $S^5$ and
the four dimensional $\mathcal{N}=4$ $U(N)$ supersymmetric
Yang-Mills (SYM) theory. The latter has a vanishing
$\beta$-function and is a conformally invariant field theory. As a
superconformal field theory the perturbative quantum corrections
appear only through the anomalous dimensions of operators. Hence
solving the field theory amounts to specifying the scaling
dimensions of the operators (up to some conformal ratios).
The scaling dimension $\Delta$ of a
given operator $\cO_\Delta$ is the eigenvalue of the dilatation
operator $\cD$, i.e.
\be\l{scaling-dimension}%
[\cD , \cO_\Delta]=\Delta \cO_\Delta\ .
\ee%
The dilatation operator $\cD$ is the Hamiltonian of the $\cN=4$
gauge theory on $\mathbb{R} \times S^3$ or equivalently the
Hamiltonian of the gauge theory on $\mathbb{R}^4$ in radial
quantization \cite{Aharony:1999ti}. Therefore, giving a
representation of $\cD$ on the space of fields of the $\cN=4$
gauge theory (the constituents of $\cO_\Delta$) amounts to solving
the theory.

In the last two years, motivated by the results and insights
obtained from the BMN \cite{BMN} double scaling limit (for reviews
see \cite{Plefka, BMNreview}) some very important steps towards
determining the dilatation operator $\cD$ were taken
\cite{Beisert:2003tq,Beisert:2003jj}. The original observation was
made realizing a close connection between the dilatation operator
$\cD$ in some specific subsector of the operators of the gauge
theory and the Hamiltonian of an integrable system, namely the
$SO(6)$ spin chain system \cite{Minahan-Zarembo}. Based on this
observation it was proposed that the $\cN=4$ gauge theory is also
integrable (in a certain limit). This
proposal is based on two facts:\\
\hspace*{.3cm} ({\it i}) There is a one-to-one correspondence
between the gauge invariant \opt s of the $\cN=4$ gauge theory
which are built strictly from $\Delta_0$ number of (six real)
scalars of the $\cN=4$ gauge multiplet and the allowed
configurations of an $SO(6)$ spin chain system with $\Delta_0$
number of sites. Verification of this observation is almost
immediate.\\
\hspace*{.3cm} ({\it ii}) At planar, one-loop level, i.e.\! strict
't Hooft large $N$ limit and at first order in the 't Hooft
coupling $g_{YM}^2 N$, the one-loop anomalous correction to the
dilatation operator obtained via explicit computations of
two-point functions matches exactly the Hamiltonian of an $SO(6)$
spin chain system with nearest neighbor interactions; for a nice
review on this subject see \cite{Beisert-review}.

The above observations were extended beyond the $SO(6)$ sector to
the one-loop planar dilatation operator of the full theory
\cite{Beisert:2003tq}, which matches the Hamiltonian of a ``super
spin chain'' system \cite{Beisert:2003yb}. The above has been
checked by explicit computations of two-point functions. In fact
it has been shown that the four-dimensional superconformal
invariance under $psu(2,2|4)$ is strong enough to completely fix
the form of the dilatation operator at one-loop planar level
\cite{Beisert-review}.

To argue for the integrability of the $\cN=4$ SYM, even in the
strict 't Hooft planar limit, one needs to know all loop
results.\footnote{For appearances of integrable structures in
string theory see  \cite{Kruczenski:2003gt,Kruczenski:2004wg} and
\cite{Bena:2003wd}.} In this direction the higher loop planar
dilatation operator has been worked out in
\cite{Beisert-higher-loops, Long-range}, where it was argued that
the integrability structure survives. As a result of technical
difficulties, these computations have been mainly limited to some
subsectors of the $SO(6)$ sector. At higher-loop level, although
still very restrictive, the superconformal symmetry is not enough
to completely fix the form of the dilatation operator
\cite{Beisert-review} and some explicit computation is necessary.
These computations, on the gauge theory side, have been performed
mainly in the BMN or near BMN limit \cite{Beisert-higher-loops,
Long-range} corresponding to the thermodynamic limit of the spin
chain system where the Bethe ansatz \cite{Bethe} is applicable
\cite{Beisert-review}. At higher-loop level the corresponding spin
chain system is not of the form of nearest neighbor interactions.
So far the dilatation operator has been worked out up to four-loop
planar level; at three-loop level some discrepancies with the
results of the string theory side have been observed
\cite{Beisert-review, higher-loop-failing} and more recently it has been argued that these 
discrepancies can be resolved using
the ``quantum string Bethe ansatz'' \cite{QBethe}.\footnote{In a
parallel line of development, the possible existence of integrable
structures in gauge theories has attracted interest, both in the
guise of self-dual Yang-Mills \cite{selfdualYM} and in the more
phenomenologically interesting QCD \cite{QCDspinchain,
QCDspinchain2}. In \cite{KL03} an integrable structure originally
found in QCD was used to compute the anomalous dimensions of
certain Wilson operators in the $\mathcal{N}=4$ gauge theory.}

In this work we consider the less supersymmetric cases of $\cN=1$
superconformal gauge theories and extend the results of the spin
chain/gauge theory correspondence to these cases. The example of
interest here is the  $\cN=1,\ D=4$ $U(N)^{p\times q}$ SYM with $3
\, p \, q$ chiral matter fields in the bi-fundamental
representations $(N_{ij},\bar N_{i+1,j})$, $(N_{ij},\bar
N_{i,j+1})$ and $(N_{i+1,j},\bar N_{i,j+1})$, with $i=1,2,\cdots,
p$, $j=1,2,\cdots, q$. This gauge theory is dual to type IIB
strings on the $1/8$ BPS orbifold of $AdS_5 \times S^5$
\cite{Douglas:1996sw,Eva-Shamit, vafa-et.al.}. We find the
dilatation operator of this SCFT and argue for the integrability
of the $\cN=1$ theory in the appropriate limit \cite{Wang:2003cu,
Chen:2004mu}, which we find to extend beyond the untwisted states
that result from large $N$ orbifold inheritance
\cite{Eva-Shamit,vafa-et.al.}. A similar analysis for $\cN=2$
superconformal gauge theories has been carried out in
\cite{Semenoff}. (Penrose limits of such orbifolds have been
studied in \cite{Compactified-ppwaves}.)

One of our remarkable results is that we find a three-dimensional
classical statistical mechanical system whose Euclidean spatial
slices form the quiver (moose) diagram \cite{Douglas:1996sw}
describing the orbifold gauge theory, which for the case of
interest is a two-dimensional triangular lattice. We will argue
that our system is equivalent to a $2+1$ dimensional $U(N)$
lattice gauge theory. This in turn is equivalent to a $2+1$
dimensional string theory with discretized worldsheet and target
space. The thermodynamic limit then corresponds to taking $N$ large.
The latter brings a new insight into the spin chains related to
the $\cN=4$ gauge theory.

In section \ref{gauge-theory-on-lattice}, we outline the structure
of the lattice, the transformation properties of the
bi-fundamental fields, enumerate the gauge invariant operators and
describe their structure on the lattice. In section
\ref{dilatation-opt}, we work out the dilatation operator of the
$\mathcal{N}=1$ orbifold theory. We demonstrate that the
dilatation \opt\ can be thought of as the transfer matrix for
 a theory on the
corresponding $2d$ lattice. In section \ref{dynamical-pictures},
we discuss two different views of the structure we uncover: a
description in terms of a lattice Laplacian which makes manifest
the string dynamics and a description in terms of an integrable
spin chain. We also make some comments on the BMN limit of the
$\cN=1$ superconformal theory. In section \ref{branebox}, we
discuss how the $2+1$-dimensional lattice picture is extended to a $3+2+1$
dimensional string theory, once the gauge fields of the $\cN=1$
theory are also included.
For this purpose we use a relation between
AdS/CFT and brane box models \cite{Hanany:1997tb}. In section
\ref{Higgs}, we discuss the Higgsed phase of the $\cN=1$ theory,
where the conformal symmetry is lost. In this section we
discuss the relation to $2+1$ dimensional lattice gauge
theories as well as the deconstruction and six-dimensional
picture. Finally in section \ref{summary} we give a summary and
outlook. In two appendices we introduce and fix our conventions,
and give some examples to clarify the discussion in the main body
of the work.

\section{$\cN=1$ Gauge Theory and the Lattice}
\label{gauge-theory-on-lattice}

In this section we introduce and elaborate on a pictorial way of
presenting the $\cN=1$  SYM theory, an interesting subset of its
operators and the dilatation \opt , using the corresponding quiver
diagram, which in our case is a two dimensional triangular
lattice. This two dimensional lattice plays a central part in our
construction, and its role as a target space for string dynamics
will  become apparent as we proceed.

\subsection{The Lattice}
\label{the-lattice}

The field content of the $\mathcal{N}=1$ supersymmetric gauge
theory is given by a triplet of chiral supermultiplets, which we
label as  $A, B, C$. These arise as orbifold projections of the
three chiral multiplets of $\mathcal{N}=4$ theory when written in
$\mathcal{N}=1$ language.\footnote{The $\mathcal{N}=4$ action,
written in $\mathcal{N}=1$ language, is presented in Appendix A.
In Appendix B we give an example of an explicit projection which
demonstrates how the transformation properties, which are
responsible for the structure of the Moose diagram, arise.} In
addition, we have a single vector supermultiplet, again projected
from the vector multiplet of the parent $\mathcal{N}=4$ theory.

The projected fields transform non-trivially under the
$U(N)^{p\times q}$ subgroup of the original $U(Npq)$ gauge
symmetry of the $\cN=4$ theory which survives the orbifolding.
This subgroup consists of the degrees of freedom left invariant by
the orbifold action. To construct the $\mathbb{Z}_p\times
\mathbb{Z}_q$ orbifold theory we start with an $\cN=4$ $U(Npq)$
theory and then find an $Npq\times Npq$ representation of
$\mathbb{Z}_p\times \mathbb{Z}_q$ under which the chiral
multiplets are projected to $3pq$ $N\times N$ matrices, in
bi-fundamental representations  of the $U(N)^{p \times q}$
\cite{Eva-Shamit, vafa-et.al.}.\footnote{For a generalization of the orbifolds to other quiver paths on a tours see  \cite{general-quiver}.} 

The three complex scalars are bi-fundamentals under this gauge
symmetry, and are to be associated with directed links on the
lattice, the indices $i,j$ denoting the particular gauge group
associated with a lattice site.

The transformation properties are as follows:
\begin{subequations} \label{transforms}
\begin{align}
  A_{i,j} &
   \: \: \: \: \: \: \: \: \sim \: \: \: \: \: \: \:
  \left( N_{i,j} \ , \ \: \bar{N}_{i,j+1} \right) \: \: , \\
  B_{i,j} &
   \: \: \: \: \: \: \: \: \sim \: \: \: \: \: \: \:
  \left( N_{i,j} \ , \ \: \bar{N}_{i-1,j-1} \right) \: \: , \\
  C_{i,j} &
   \: \: \: \: \: \: \: \: \sim \: \: \: \: \: \: \: \:
  \left( N_{i,j} \  , \ \bar{N}_{i+1,j}  \right) \: \: .
\end{align}
\end{subequations}
The first entry gives the starting point of the directed link and
the second entry the endpoint. Conjugation of a field corresponds
to flipping the direction of the arrow on a link, yielding the
transformation properties for the conjugate fields:
\begin{subequations} \label{conjugate-transforms}
\begin{align}
  \bar{A}_{i,j} &
   \: \: \: \: \: \: \: \: \sim \: \: \: \: \: \: \:
  \left( N_{i,j+1} \ , \ \ \ \: \bar{N}_{i,j} \right) \: \: , \\
  \bar{B}_{i,j} &
   \: \: \: \: \: \: \: \: \sim \: \: \: \: \: \: \:
  \left( N_{i-1,j-1} \ , \ \bar{N}_{i,j} \right) \: \: , \\
  \bar{C}_{i,j} &
   \: \: \: \: \: \: \: \: \sim \: \: \: \: \: \: \:
  \left( N_{i+1,j} \ , \ \ \ \: \bar{N}_{i,j} \right) \: \: .
\end{align}
\end{subequations}

On the lattice, the fields in the vector multiplet $V$ of the
$\neqone$ theory are associated with lattice sites, as they are
adjoints whose transformation property is:
\begin{equation} \label{vector-transforms}
  V_{i,j}
   \: \: \: \: \: \: \: \: \sim \: \: \: \: \: \: \:
  \left( N_{i,j} \  , \ \bar{N}_{i,j}  \right) \: \: .
\end{equation}

\begin{figure}[ht]
\centering
\epsfig{figure=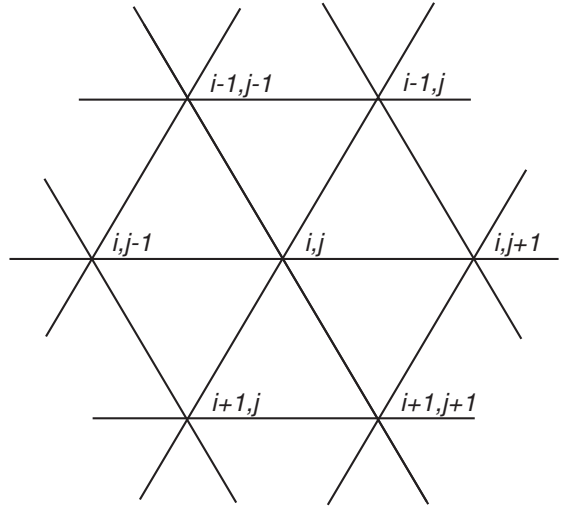, width=208pt, height=192pt}
\begin{center}
\caption{The lattice coordinates.}
\label{triangle}
\end{center}
\end{figure}

\begin{figure}[tbph]
\centering
\epsfig{figure=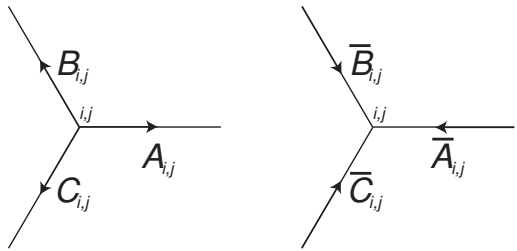,width=192pt,height=92.5pt}
\begin{center}
\caption{The direction vectors designating the three chiral
multiplets on the lattice, relative to the site $i,j$.}
\label{basis}
\end{center}
\end{figure}

The gauge structure of the theory after orbifolding is captured
succinctly by a moose (or quiver) diagram \cite{Douglas:1996sw},
whose structure gives a visualization of the transformation
properties of the fields which survive the orbifold, as given
above. For the orbifold under consideration the relevant moose is
a two-dimensional triangular lattice, with $p \times q$ sites. The
coordinates on the lattice as well as the basis vectors are
depicted in Figures \ref{triangle}
and \ref{basis}.%

In other words, our two dimensional lattice is on a torus and the
volume of the torus is proportional to $pq$; that is a {\it fuzzy}
two torus with the fuzziness parameter $\Theta=\frac{r}{pq}$,
where $r={\rm gcd}(p,q)$ is the greatest common divisor of $p$ and
$q$; for $p=q$, $\Theta=1/p$. One should, however, note that the
 lattice spacing and the size of the links is an arbitrary
scale in the conformal field theory. For this reason we have
called this lattice a ``conformal moose'' (recall the expression
in the title). This means that for the fuzzy torus we are
considering the complex structure is fixed and the ratio of the
real and imaginary parts of the Kahler form, $\Theta$, is also
fixed. In section \ref{Higgs} we consider a case where the size of
the lattice spacing is fixed, by giving VEV's to the
bi-fundamental scalars.

The ratio of the coupling of the orbifolded gauge theory to the
parent theory depends on the order of the finite group generating
the orbifold, in our case $\mathbb{Z}_p \times \mathbb{Z}_q$.
After orbifolding, the coupling of the $\mathcal{N}=1$ theory is
related to the parent theory's coupling by%
\be\label{N=1-coupling}
  g_{YM}^2 = (g_{YM}^{parent})^2 \times p \times q
\ee%
where the order of the orbifold group appears on the right hand
side.

The superpotential of the theory can also be calculated using e.g.
the techniques of \cite{Hanany:1997tb}:%
\be\label{superpotential}%
W=\sum_{i=1}^p\sum_{j=1}^q\
tr\left(A_{ij}B_{i,j+1}C_{i-1,j}-A_{ij}C_{i,j+1}B_{i+1,j+1}\right),
\ee%
where the $tr$ is over the $N\times N$ matrices and $A, B$ and
$C$'s now represent the three bi-fundamental chiral superfields.
This superpotential has a simple representation on
the $2d$ oriented triangle lattice: sum over all the basic cells
on the lattice (triangles), with a charge assignment, positive
sign to counter-clockwise loops and negative sign to clockwise
ones.

\subsection{Gauge Invariant Operators}
\label{operators}

The observables of the $\neqone$ theory we focus on are
constructed as gauge invariant combinations of the degrees of
freedom.  We shall be primarily interested in those operators
built from the three complex scalar fields which in $\neqone$
language correspond to the scalar components of the three chiral
superfields. The group structure of the fields will be responsible
for much of the interesting physics we discuss, and so apply with
little change to the fermions in the chiral multiplets. We also
briefly mention how the vector multiplet enters the picture.

Given the transformation properties of the fields, we are not free
to multiply them arbitrarily. Any gauge invariant operator is the
trace of an appropriate combination of fields, or the product of
several such traces. As is evident, any gauge invariant operator
is mapped onto a certain closed loop on the lattice, one loop
being associated to each trace. Among the closed loop operators
there exist distinguished classes, for example those which are
BPS, or those constructed only from chiral fields $A, B, C$, and
not their conjugates. We shall refer to the latter as
``holomorphic'' operators, which include as a subclass the BPS
operators. We caution the reader that not all ``holomorphic''
operators as defined are protected, and to avoid confusion, we
shall take care to distinguish between ``BPS'' and
``holomorphic''. We will show in the next section that the BPS
protected operators are those built only from a single chiral or
anti-chiral multiplet, for example a string of $A$'s, and by
virtue of gauge invariance must completely wrap the lattice in one
direction. They can of course wrap multiple times; the important
point is that they begin and end at the same site.

The most general operator consisting only of scalar fields
contains both chiral and anti-chiral fields. The chiral fields are
assigned with positive $R$-charge, whereas the anti-chiral ones
carry opposite or negative $R$-charge, as per usual in
$\mathcal{N}=1$ supersymmetry. We use the convention that the
$R$-charge of a fundamental chiral field is one. The $U(1)$
R-symmetry of the orbifold theory is a subgroup of the
$SO(6)_R\simeq SU(4)_R$ of the parent theory. In fact the subgroup
surviving the orbifolding is larger, i.e. $U(1)\times \mathbb{Z}_p
\times \mathbb{Z}_q$, where is $\mathbb{Z}_p \times \mathbb{Z}_q$
is known as ``quantum symmetry'' \cite{vafa-et.al.} and on the
lattice is nothing but the translations on the (fuzzy) two torus.
As we will see later, when we reinterpret the dilatation operator
as a Hamiltonian, the $R$-charge is a conserved quantity.

There is a straightforward way to understand which operators
survive the orbifolding. The operators we have described fall into
two classes, depending on whether they are inherited from the
parent $\mathcal{N}=4$ theory or not, i.e.\!\! untwisted or
twisted respectively. Given an operator in the daughter theory,
the question of whether the operator is inherited can be recast in
terms of its charge under the quantum symmetry.
 The generators of the $\mathbb{Z}_p \times
\mathbb{Z}_q$ quantum symmetry generate translations along the two
dimensional lattice. On the covering space the generators of
$U(N)$ are tensored with the generators of $\mathbb{Z}_p \times
\mathbb{Z}_p$ (taking $q=p$) and the gauge-invariant operators are
traces of products of fields sitting in the product representation
$U(N) \otimes \Gamma_i \otimes \tilde{\Gamma}_j$ with $\Gamma_i$
and $\tilde{\Gamma}_j$ the generators of the first and second
$\mathbb{Z}_p$ factor respectively, and $i,j=1,2,\cdots , p$. The
traces from which gauge-invariant operators are built are
understood as acting on these tensor products. Not all such
products have non-vanishing trace, and there are many
gauge-invariant operators in the parent theory which are projected
out by the orbifold action. Those which are traceless are
precisely the operators which do not survive the orbifolding.
Examples of such operators (for the case of a $\mathbb{Z}_3 \times
\mathbb{Z}_3$ orbifold) are $Tr(A^2)$, $Tr(AB)$, and $Tr(ABA)$,
with $A,B,C,\bar{A},\bar{B},\bar{C},$ the fields of
$\mathcal{N}=4$  $U(Npq)$ supersymmetric theory written as three
complex fields (as we do when writing the action for this theory
in $\mathcal{N}=1$ language).\footnote{ In our conventions we use
$Tr$ to denote trace over the $U(Npq)$ matrices of the parent
theory and $tr$ for the $N\times N$ matrices of the daughter
theory.} Some examples of operators which survive the projection
are $Tr(A^3)$, $Tr(A^6)$, $Tr(ABC)$, and $Tr(ABABAB)$. Take for
example $Tr(ABC)$. Carrying out the
projection, this gives rise to%
\[%
  \sum_{i,j} \ tr \left( A_{i,j} B_{i,j+1} C_{i-1,j} \right)
\]%
where $tr$ is now the trace over $U(N)$ valued fields. The sum in this
expression places one such operator starting at each lattice site,
and hence this operator covers the entire lattice in a symmetric
manner. It is invariant under shifts along any lattice direction
(the quantum symmetry) and the operators which are invariant under
this symmetry are precisely the inherited untwisted operators. The
Hamiltonian of the $\mathcal{N}=1$ superconformal Yang-Mills
theory, its superpotential \eqref{superpotential},  as well as the
dilatation operator are in this sector of the theory. Likewise
there are operators which appear in the $\mathcal{N}=1$ projected
theory which are not inherited from the parent. This second class
constitutes the operators forming the twisted sector of the
theory. The twisted operators are those which are not invariant
under the quantum symmetry (lattice translations), so for example
$tr \left( A_{i,j} B_{i,j+1} C_{i-1,j} \right)$ without a sum on
$i,j$, sits in the twisted sector. It is also evident that not all
BPS operators in the daughter theory are descendants of chiral
primary operators in the parent $\mathcal{N}=4$ theory. An
operator of the form $tr(A_{i,j} A_{i,j+1} \ldots A_{i,j+p})$
which wraps the lattice once along a single horizontal line is not
inherited, but if we replicate it along all horizontal lines,
$\sum_{i} tr(A_{i,j} A_{i,j+1} \ldots A_{i,j+p})$, then it can be
related to an $\mathcal{N}=4$ chiral primary.

There is another classification of the \opt s on the lattice which
comes from the fact that the lattice is on a torus. As we argued
above the gauge invariant \opt s are orientable close loops on the
lattice, they can then be shrinkable or non-shrinkable cycles of
the (fuzzy) two torus. For example, the \opt\ $tr \left( A_{i,j}
B_{i,j+1} C_{i-1,j} \right)$ is shrinkable whereas $tr(A_{i,j}
A_{i,j+1} \ldots A_{i,j+p})$ is a non-shrinkable one. One can
associate a ``winding'' number to non-shrinkable \opt s. We will
comment more on this point in section \ref{dynamical-pictures}.

We would like to also point out that, in the 2+1-dimensional picture
given here,
one may introduce
Wilson lines to generate gauge invariant operators that cover
all six dimensions, by extending our loops to also include
loops that extend into the 3+1-dimensional space-time. The idea
is to take a composite operator, sitting at $x$, and explode it into
pieces sitting at different space-time points, with Wilson lines
running from space-time point to space-time point between the
sub-operators at the
different points to make the whole operator gauge invariant. Then
we have a loop that lies in 3+1+2 dimensions. The
dilatation/Hamiltonian operator is then a direct sum of the 3+1-dimensional
one and the one on the lattice (to be discussed in the next section).

\section{Gauge Theory Dynamics on the Lattice}
\label{dilatation-opt}

So far we have built a one-to-one relation between the gauge
invariant \opt s of a $U(N)^{p\times q}$ $\cN=1$ superconformal
gauge theory which are made out of $3pq$ bi-fundamental scalars
and the oriented closed loops on the $2d$ triangle lattice, which
wraps a fuzzy two torus. Here we extend the lattice description of
the $\cN=1$ SCFT to beyond the level of \opt s and Hilbert spaces,
to a dynamical level and in section \ref{transfer-matrix} present
a simple lattice description of the terms in the one loop planar
dilatation operator.

\subsection{Dilatation operator of the ${\cal N}=1$ gauge theory}
\label{gauge-theory-orbifold}


In this section we work out the dilatation operator of the
$\mathcal{N}=1$ superconformal gauge theory, in the subsector of
operators built purely from the three complex scalars. We present
a derivation via an orbifolding of the $\mathcal{N}=4$ dilatation
operator, but also outline a more direct derivation starting from
the $\mathcal{N}=1$ action.


The $\mathcal{N}=4$ dilatation operator up to two-loop order and
at planar-level for operators built strictly from scalars (and no
covariant derivatives) has been worked out in
\cite{Beisert:2003tq}. We quote the one-loop result here, which we
take as a starting point for the derivation of the dilatation
operator in the $\mathcal{N}=1$ theory. This theory contains six
real scalars transforming in the adjoint representation of the
$U(Npq)$ gauge group and the fundamental of the $SO(6)$ R-symmetry
of the $\mathcal{N}=4$ theory, which we collect into three complex
scalars making manifest an $SU(3)$ subgroup of $SO(6)$.

The dilatation operator $\mathfrak{D}$ has a perturbative
expansion in powers of the Yang-Mills coupling constant $g_{YM}$,
taking the form
\begin{equation} \label{extract-coupling}
  \mathfrak{D} = \sum_{n=0}^\infty
  \left( \frac{g_{YM}}{4 \pi} \right)^{2n} \mathfrak{D}_{n} \: \: .
\end{equation}%
Here $\mathfrak{D}_{n}$ is the $n$-loop contribution. The
eigenvectors of $\mathfrak{D}$ are the operators with well defined
scaling dimensions, given by their respective eigenvalues.  In
general at higher-loops $\mathfrak{D}$ is not diagonal due to
operator mixing, and finding a suitable basis of scaling operators
requires diagonalizing $\mathfrak{D}$ at the appropriate loop
order. The tree level (classical) scaling dimension is
$\mathfrak{D}_{0}$, which is given by\footnote{See Appendix A for
conventions and definitions.}
\begin{equation}
  \mathfrak{D}_0 = \sum_{m=1}^6 Tr : \phi_m \ \delta_m : \: \: ,
\end{equation}
where $: \ :$ denotes normal ordering, taken here to mean that all
the variations with respect to the fields are understood not to
act on other fields within the same $: \ :$ block.

The one-loop correction to the scaling is given by (after
extracting the coupling dependent prefactor)
\begin{equation} \label{dilatation-real}
  \mathfrak{D}_1 \: = \: -
  \sum_{m=1}^6 \sum_{n=1}^6 Tr : \left(
  [ \phi_m , \phi_n ]
  [ \delta_m , \delta_n ]
  +
  \frac{1}{2}
  [ \phi_m , \delta_n ]
  [ \phi_m , \delta_n ]
  \right) : \: \: .
\end{equation}
The ordering of the fields is important because of their matrix
structure. As explained in Appendix A, the derivatives which
appear above are a short-hand way of capturing the action of Wick
contractions, leading to propagators.

The one-loop dilatation operator, when expressed in terms of the
complex scalars, can be split into three parts,%
\be \label{n4-dilatation-full}
  \mathfrak{D}_1 \: = \:
  \mathfrak{D}_1^h \: + \: \mathfrak{D}_1^{\bar{h}} \: + \: \mathfrak{D}_1^{h \bar{h}} \: \: ,
\ee%
with%
\be \label{n4-dilatation}
\begin{split}
  \mathfrak{D}_1^h =
  & -
    \sum_{i=1}^3 \sum_{j=1}^3
  Tr \Big(
  [\Phi_i,\Phi_j][\Delta_i,\Delta_j]
  \Big) \: \: , \\
  \mathfrak{D}_1^{\bar{h}} =
  & -
  \sum_{i=1}^3 \sum_{j=1}^3
  Tr \Big(
  [\bar\Phi_i,\bar\Phi_j][\bar\Delta_i,\bar\Delta_j]
  \Big) \: \: , \\
  \mathfrak{D}_1^{h \bar{h}} =
  & -
  \sum_{i=1}^3 \sum_{j=1}^3
  Tr \Big(
  2 [\Phi_i,\bar\Phi_j][\Delta_i,\bar\Delta_j] +
  [\Phi_i,\Delta_j][\bar\Phi_i,\bar\Delta_j] +
  [\Phi_i,\bar\Delta_j][\bar\Phi_i,\Delta_j]
  \Big) \: \: .
\end{split}
\ee%
The significance of explicitly splitting the dilatation operator
in this way will become clear momentarily. Here $\mathfrak{D}_1^h$
denotes the parts of the dilatation operator constructed only from
holomorphic fields and derivatives, and likewise
$\mathfrak{D}_1^{\bar{h}}$ with holomorphic fields and derivatives
replaced by anti-holomorphic ones. Finally, $\mathfrak{D}_1^{h
\bar{h}}$ contains mixed holomorphic and anti-holomorphic terms.
Certain subsectors of operators are not  mixed under
renormalization; the example of our interest is the sector of
holomorphic operators which is closed. This is due to the fact
that $\mathfrak{D}_1^{\bar{h}}$ and $\mathfrak{D}_1^{h\bar{h}}$
vanishes on the holomorphic operators and the action of the
dilatation operator on holomorphic operators receives
contributions only from $\mathfrak{D}_1^h$.

To obtain the dilatation \opt\ of the $\cN=1$ theory we perform
the orbifolding on \eqref{n4-dilatation}. The justification why
this procedure should work comes from the fact that the dilatation
\opt\ of the $\cN=1$ theory is in the {\it untwisted} sector and
hence is directly inherited from the parent theory
\cite{vafa-et.al.}. To perform the orbifolding, we expand the
fields and the variations in equation \eqref{n4-dilatation} in a
basis of the orbifolded generators $\Phi_I = \Phi_I^a T^a,
\Delta_I = \Delta_I^a \bar{T}^a$ (as explained in Appendix B; see
also Appendix A for why $\Delta_I$ is taken to transform as the
conjugate of $\Phi_I$), then collect terms after evaluating the
trace. Here $a$ enumerates the hermitian  generators of $U(Npq)$.

Carrying out the projection for a general $\mathbb{Z}_p \times
\mathbb{Z}_p$ orbifold (i.e.\! taking $q=p$), we arrive at a sum
of terms, whose structure is best captured in terms of interaction
``plaquettes'' on a ``Moose'' or ``quiver'' lattice. This is
described in detail in the next section, where we also
re-interpret the dilatation operator as a Hamiltonian or transfer
matrix for a certain lattice theory. This is in line with the
recent philosophy pursued in the spin chain constructions for the
$\mathcal{N}=4$ super Yang-Mills theory, where this picture has
led to insights about integrability of the $\mathcal{N}=4$ theory
in certain regimes, and has allowed the use of techniques such as
the Bethe Ansatz for finding a diagonal basis of scaling
operators. Our construction in the next section brings out
interesting dynamics not previously noted in the $\mathcal{N}=4$
studies.

Having obtained the dilatation \opt\ a few comments are in
order.\\
First, the $U(1)$ R-charge is a conserved quantity. This property
is a direct consequence of the fact that every term in the
$\mathcal{N}=4$ interaction Hamiltonian carries zero net
$R$-charge, implying the same for every term in the dilatation
operator \eqref{n4-dilatation-full}. Together with the vanishing
$R$-charge of the vacuum, this means that the two-point
correlation functions of operators with different $R$-charges
automatically vanishes, and hence the dilatation operator won't
connect operators of different $R$-charge. The plaquettes we
construct in the next section, which correspond term by term to
the dilatation operator, also reflect this fact.\\
Next, as explained above and can be readily seen from
\eqref{n4-dilatation}, the holomorphic \opt s (like-wise for
anti-holomorphic \opt s) form a closed sector under the action of
the dilatation \opt\ and this is the sector which we will mainly
focus on in this paper. For holomorphic operators, the classical
(engineering) dimension is equal to the total $R$-charge of the
operator, and this dimension is also the length of the loop on the
quiver lattice, measured in units of lattice length (which due to
the conformal invariance is an arbitrary length scale).
 Conservation of
$R$-charge then implies conservation of dimension and as well as
lengths of loops on the lattice when the loops are allowed to
evolve. This conservation extends to non-planar string joining and
splitting interactions. Of course, conservation of the length
applies generally to all operators made out of bi-fundamentals,
being a result  of the fact that in every term of the dilatation
operator \eqref{n4-dilatation-full}, the number of fields and
derivatives are the same.\footnote{In our discussions we mainly
focus on operators built from bi-fundamental fields. If we include
fields in the vector multiplet in our operators, the equality of
dimension and length no longer holds. As pointed out in the
previous section, fields in the vector multiplet transform as
adjoints, and hence sit at sites on the lattice, not on links. As
such, they do not contribute to the length of the loops on the
lattice, but do of course affect the dimension. In the dilatation
operator  constructed in \eqref{n4-dilatation-full} we have
assumed  the absence of the vector multiplet terms, and our
discussion of string dynamics below is special to this case.
Inclusion of the gauge fields can be done in a similar way
starting with the full one-loop planar dilatation \opt\ of the
$\cN=4$ theory \cite{Beisert:2003jj}. We will comment on the
inclusion of the vector multiplets section \ref{branebox}.} The
identification of the $R$-charge and dimension is, however,
special to holomorphic and anti-holomorphic operators.

Conservation of the length is the statement that the two-point
functions of two renormalized operators is non-vanishing only when
they carry the same classical dimension. The area enclosed by the
loops, however, is not conserved. As we will see the area is
restricted to change by the area enclosed by zero or two
fundamental lattice triangles at each Euclidean time step. The
behavior of the anti-holomorphic operators exactly mirrors that of
the holomorphic operators, so we restrict our attention to the
holomorphic ones.

Another route to the derivation of the $\mathcal{N}=1$ dilatation
operator exists, in which we start with the explicit form of the
$\mathcal{N}=1$ action. The derivation of the $\mathcal{N}=1$
action itself progresses via an orbifolding of the known
$\mathcal{N}=4$ theory. Then, using standard Feynman diagram
techniques we compute two-point functions of composite operators.
As usual, correlators of local composite operators require
renormalization beyond those necessary for fundamental fields
appearing in the action, and introduce anomalous dimensions for
the composite operators (see for example \cite{BMNreview}). From
the renormalized two-point functions of such operators, we then
extract their scaling dimensions, which  defines the dilatation
operator \cite{Beisert:2003tq}.

The two approaches differ in the order in which the orbifolding is
applied and the dilatation operator computed. As presented in this
section, the dilatation operator of the $\mathcal{N}=4$ is first
constructed, to which the orbifolding is applied, yielding the
dilatation operator of the $\mathcal{N}=1$ theory. The alternative
of first orbifolding the $\mathcal{N}=4$ theory prior to using it
to derive the dilatation operator produces the same result, since
the orbifold projection as applied to operators appearing in the
$\mathcal{N}=4$ Hamiltonian is precisely the same as the one
applied to the dilatation operator, and the one- and higher-loop
structure of the dilatation operator is intimately tied to the
structure of the Hamiltonian. As a result, the orbifolding
commutes with the action of the dilatation operator. For operators
in the untwisted sector, this is required by orbifolding
inheritance, which applies at the planar level we consider, but
the result is in fact more general. The fact that orbifolding does
not destroy the structure of the dilatation operator can be shown
noting that the dilatation \opt\ of $\cN=4$ SYM commutes with the
$SO(6)$ $R$-symmetry generators and hence with the
$\mathbb{Z}_p\times \mathbb{Z}_q\subset SO(6)$ with respect to
which we do the orbifolding.

A similar construction to the one we use here has been presented
in \cite{Semenoff}, where an $\mathcal{N}=2$ supersymmetric theory
is derived from the $\mathcal{N}=4$ theory via a $\mathbb{Z}_p$
orbifolding. Our dilatation operator is related to that of
\cite{Semenoff} by a second $\mathbb{Z}_q$ projection.  Our
approach in this section is quite general, and can be applied
generically to other orbifolds.
\subsection{The Time Evolution Matrix and Interactions}
\label{transfer-matrix}

The dilatation operator in the $\cN=1$ superconformal Yang-Mills
gauge theory on $R^4$ is the Hamiltonian in the radial
quantization and/or the Hamiltonian of the gauge theory on
$R_\tau\times S^3$.\footnote{From the gauge theory on
$R_\tau\times S^3$ viewpoint one can of course trivially move
between the Euclidean and Minkowski pictures. For our purposes we
prefer to work with the Euclidean time.} Next, recall that \opt s
of different classical dimension cannot be related by $\cD$. That
is, if $\cD_0 \cO_1=d_1\cO_1$, $\cD_0 \cO_2=d_2\cO_2$ and $d_1\neq
d_2$, then $\langle \cO_1|\cD|\cO_2\rangle =0$. Hence, the
classical dimension of operators is a conserved quantum number
under the time evolution generated by the dilatation operator.
Therefore, we can easily remove the $\cD_0$ part in the dilatation
operator $\cD$ and if we restrict ourselves to the one-loop planar
dilatation operator, $\cD_1$, we have:%
 \be\label{time-evolution}%
(1+\epsilon \cD_1)\cO_\tau=\cO_{\tau+\epsilon}\ .%
\ee%
 In other
words, $\cD_1$ may be thought as the operator evolving the
configuration at ``time'' $\tau$ to a configuration at time $\tau
+\epsilon$, with $\epsilon$ the minimal discrete time step. Since
the theory is conformal there is no preferred scale in the theory,
neither for $\epsilon$ nor for the triangle lattice spacing and we
can then smoothly take the $\epsilon\to 0$ limit.

In the lattice theory terminology an operator which generates
transitions between the configurations in different time steps is
called a {\it transfer matrix}. We identify the matrix elements of
the one-loop planar dilatation operator $\mathfrak{D}_1$ between
the basis of states given by the gauge-invariant operators with
the matrix elements of the transfer matrix $\hat{T}$ in the same
basis, i.e.,%
 \be\label{Tmatrix-D1}%
  \big< \: \mathcal{O}_i \: | \: \hat{T} \: | \: \mathcal{O}_j \: \big> \: \equiv \:
  \big< \: \mathcal{O}_i \: | \: \mathfrak{D} \: | \: \mathcal{O}_j \: \big> \ ,
\ee%
with $\mathcal{O}_i$ labeling the basis of gauge-invariant
operators, consisting of any number of traces. The finite
translations in time can then be obtained iterating the action of
the transfer matrix, i.e. the transfer matrix to power
$T/\epsilon$, in the $\epsilon\to 0$ limit, produces the
translation by finite amount $T$. We focus our discussion on
single trace operators, except when we discuss $1/N$ corrections
which lead to string joining or splitting, as the more general
cases of multiple-trace operators follows immediately from their
study. The transfer matrix for our system is then an infinite
dimensional matrix, since there are an infinite number of
gauge-invariant operators, of arbitrary dimension, that can be
specified at each slice, even on a finite size lattice (since we
allow our operators to wrap the lattice any number of times.) Note
that the transfer matrix is block diagonal for certain subclasses
of operators. The rows and columns labeled by BPS operators have
vanishing entries at the one-loop level, as the overlap of a BPS
operator vanishes with all operators (including itself). This is
what defines them as BPS. Had we included the tree-level
contribution, for BPS operators only the diagonal elements would
be non-zero.\footnote{In this section we identify the transfer
matrix with the one-loop dilatation operator. We could have
included the tree-level contribution as well, as we will do in
section \ref{spin-chain}. This leads to a trivial change in the
present discussion, and we drop the tree-level contribution for
now in the interest of clarity. We will reintroduce it when
necessary in section \ref{spin-chain}.}

Another set of blocks is formed by the holomorphic (likewise
anti-holomorphic) operators, since as mentioned in section
\ref{gauge-theory-orbifold}, they will not mix under
renormalization with operators outside the class. This is
basically due to the fact that $\cD_1$ is a sum of three parts
$\cD_1^h$, $\cD_1^{{\bar h}}$ and $\cD_1^{h{\bar h}}$, and that
$\cD_1^{{\bar h}}$ and $\cD_1^{h{\bar h}}$ have derivatives with
respect to anti-holomorphic fields ({\it cf.}
\eqref{n4-dilatation}). These statements are predicated on a
specific choice of transfer matrix, which defines our statistical
mechanical system, and which we now specify.

\begin{figure}[ht]
\centering
\epsfig{figure=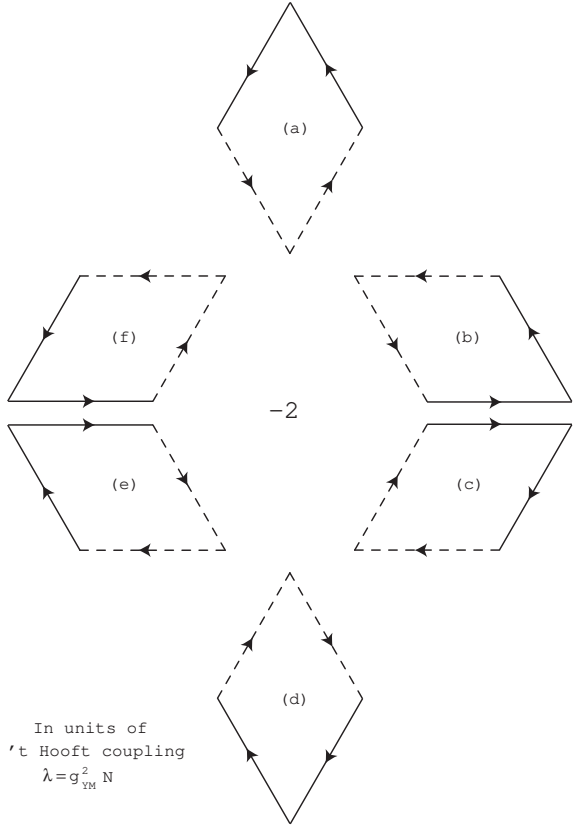,width=214.269pt,height=311.99pt}
\begin{center}
\caption{The basic ``holomorphic'' interaction plaquettes which
enclose two units of lattice area. The solid lines correspond to
fields, and the dashed lines to field derivatives. These are the
terms appearing in $\mathfrak{D}_1^h$, and contribute with a
coefficient of $-2$ (in units of $g^2_{YM}N$) to the amplitude.
Notice that the bottom three are mirrors of the top three, flipped
along the horizontal.} \label{open-plaquettes}
\end{center}
\end{figure}

\begin{figure}[th]
\centering
\epsfig{figure=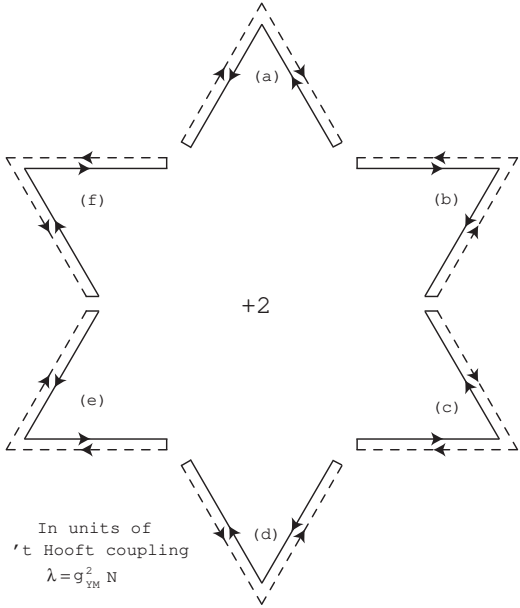, width=194.3pt,height=228.745pt}
\begin{center}
\caption{The basic ``holomorphic'' interaction plaquettes which
enclose zero area. As before, the solid lines correspond to
fields, and the dashed lines to field derivatives. These are the
terms appearing in $\mathfrak{D}_1^h$, and contribute $+2$ (in
units of $g^2_{YM}N$) to the amplitude. These terms are related
individually to those in Figure \ref{open-plaquettes} by commuting
the derivative terms, as they appear in $\mathfrak{D}_1^h$. Again,
the bottom three are mirrors of the top three, flipped along the
horizontal.} \label{corners}
\end{center}
\end{figure}

The evolution of the operators from time slice to time slice gives
rise to a natural picture in terms of fluctuating strings and
their interactions. In section \ref{operators} we have prescribed
a map from local gauge-invariant operators to loops on the quiver
diagram. This quiver diagram plays the role of a lattice and the
loops behave as discretized strings propagating in time. The
allowed fluctuations of these strings are determined by the
structure of the terms in the transfer matrix. We can describe
these allowed transitions in terms of ``interaction plaquettes''.
Some examples will clarify the picture. For concreteness, first
consider the  holomorphic interactions, i.e.\! those plaquette
terms appearing in $\mathfrak{D}_1^h$. There are two basic types
of plaquettes to consider, those which enclose two unit triangle
areas and those which enclose zero area (the open plaquettes and
corners).  They are shown in Figures \ref{open-plaquettes} and
\ref{corners} respectively. Notice that in Figure
\ref{open-plaquettes} the three plaquettes in the top half are
traversed in a right-handed fashion, and the bottom half in the
left-handed direction.

\begin{figure}[ht]
\centering
\epsfig{figure=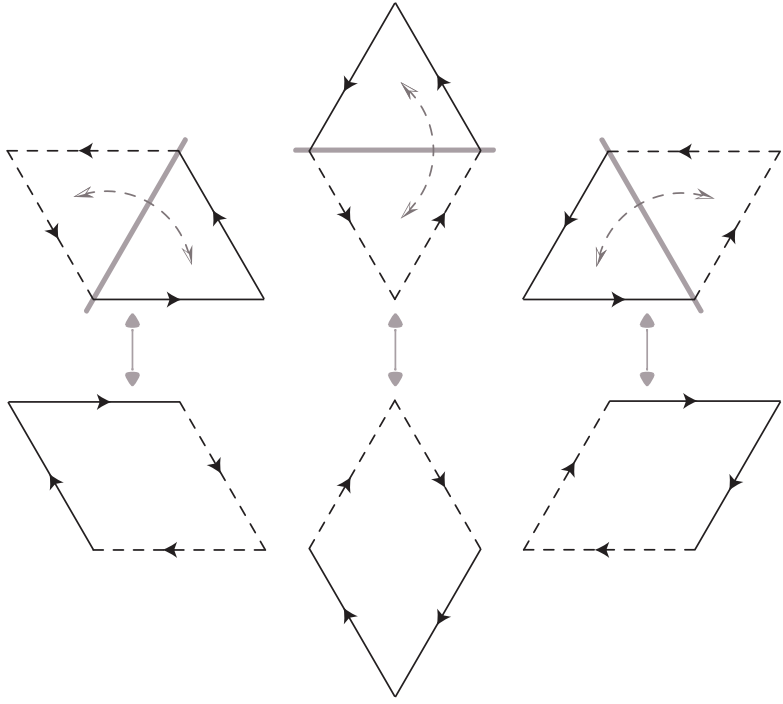, width=293.51pt,height=263.74pt}
\begin{center}
\caption{The diagrammatic relation between left-handed and
right-handed open plaquettes.}
\label{transform1}
\end{center}
\end{figure}

In Figure \ref{transform1} we show how the right-handed
parallelogram plaquettes are related to the left-handed ones.
Consider an operator of the form $tr(A B \Delta^A \Delta^B)$
(plaquette (b) in Figure \ref{open-plaquettes}).\footnote{ In
taking the traces one must also take care in placing the correct
lattice site indices on the fields and derivatives appearing in
these operators. Here our notation for operators presupposes such
indices. When reversing the trace, we supply new indices as needed
to make the operator properly gauge-invariant.} Reading the trace
in the opposite direction, we get $tr(B A \Delta^B \Delta^A)$,
transforming a right-handed operator into a left-handed one. These
both arise from $\mathfrak{D}_1^h \sim
Tr([A,B][\Delta^A,\Delta^B]) + \ldots$, which gives rise to a sum
of terms of the form $tr(A B \Delta^A \Delta^B)$ and $tr(B A
\Delta^B \Delta^A)$ at all lattice sites, and so the left-right
handed flip arises from the commutation of $A,B$ and
$\Delta^A,\Delta^B$ in $\mathfrak{D}_1^h$.

\begin{figure}[ht]
\centering
\epsfig{figure=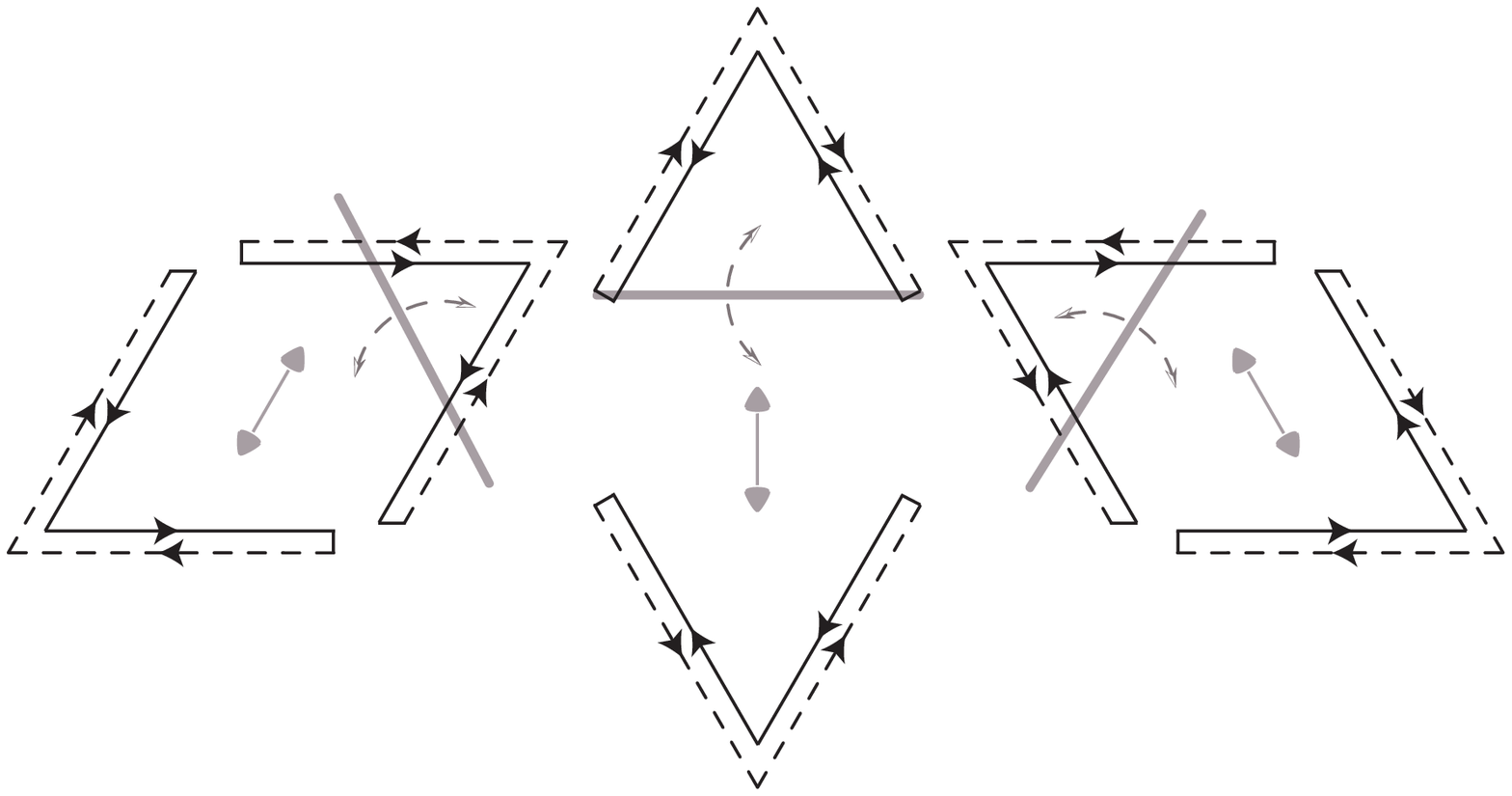, width=331.46pt,height=172.53pt}
\begin{center}
\caption{The diagrammatic relation between left-handed and right-handed corner plaquettes.}
\label{transform2}
\end{center}
\end{figure}

Diagrammatically, this reversal corresponds to flipping the
plaquette along the diagonal dividing the plaquette into two
equilateral triangles, which are fundamental units of the lattice
(this diagonal also separates the fields and the derivatives).
Similarly, Figure \ref{transform2} shows how the right-handed and
left-handed corner plaquettes are related.
\begin{figure}[ht]
\centering
\epsfig{figure=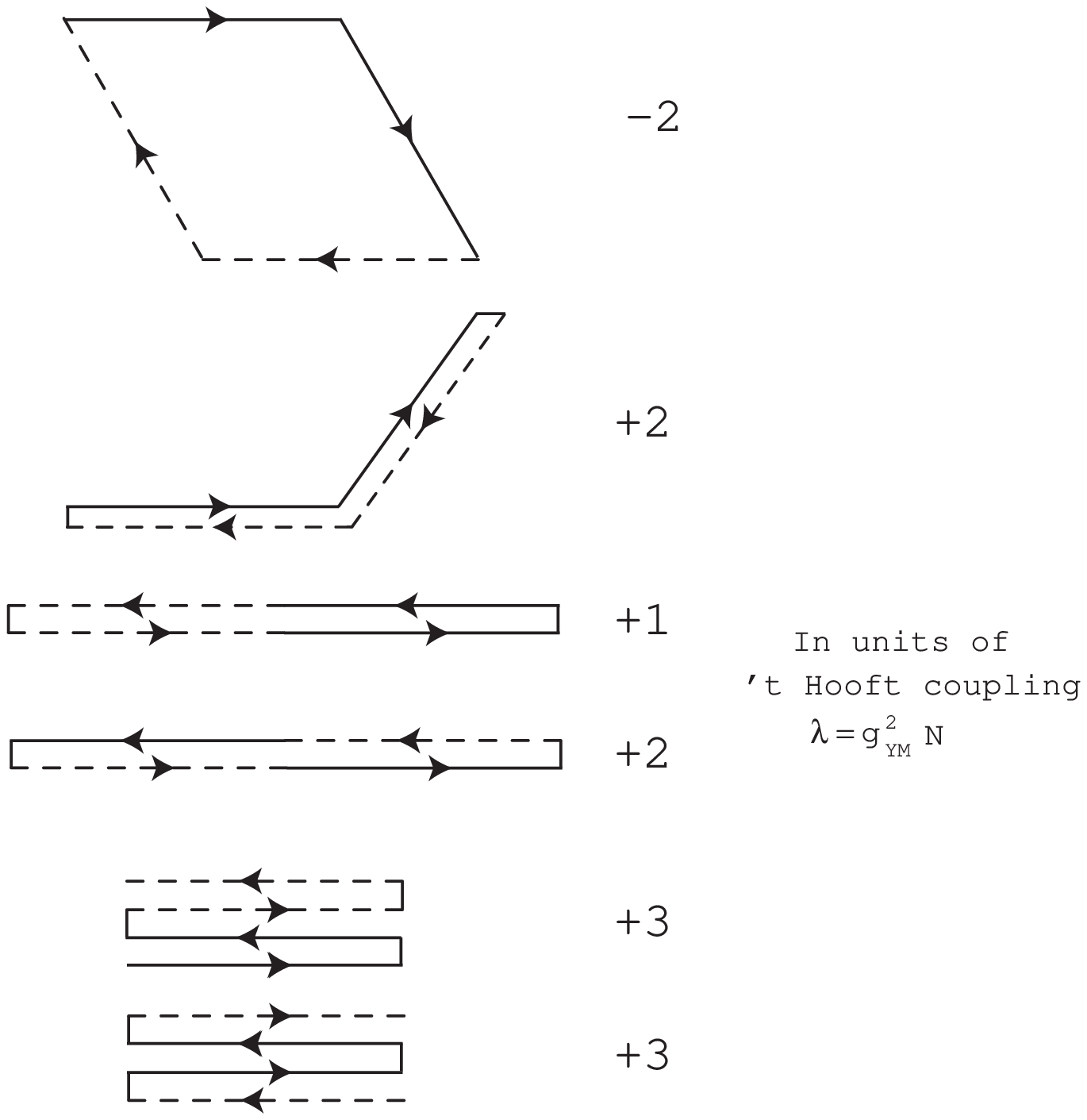, width=237.7pt,height=260.975pt}
\begin{center}
\caption{Examples of operators which mix holomorphic and anti-holomorphic
fields. The open plaquette has the same weight as in Figure
\ref{open-plaquettes}, and the corner plaquette has the same weight as in
Figure \ref{corners}. Due to some cancelations the third diagram has half
the weight of a corner plaquette in Figure \ref{corners}, while the
fourth diagram carries the same weight as the corner plaquettes.
The final two figures carry the weight $+3$. The numbers on the right side of
the plaquettes shows their contributions in units of $g^2_{YM}N$.}
\label{non-hol}
\end{center}
\end{figure}
Hermitian Conjugation of a plaquette simply reverses the direction in which the arrows flow. For example, conjugation of $tr(A \bar{C} \Delta^A \bar{\Delta}^C)$ gives
$tr(C \bar{A} \Delta^C \bar{\Delta}^A)$, turning the original right-handed plaquette into
a left-handed one. Each plaquette in $\mathfrak{D}_1^h$ is the conjugate of one plaquette in
$\mathfrak{D}_1^{\bar{h}}$. The sum of $\mathfrak{D}_1^h$ and $\mathfrak{D}_1^{\bar{h}}$ is hermitian.
Likewise, the first term in $\mathfrak{D}_1^{h \bar{h}}$ is hermitian and the sum of the second and third terms is also hermitian. Thus $\mathfrak{D}_1$ is hermitian with real eigenvalues, as expected, since it is the one-loop correction to the generator of the conformal group of the theory, and whose eigenvalues, giving the one-loop corrections to the scaling dimensions, must be real.

When expanding the sum in $\mathfrak{D}_1$, we have terms such as
\be \label{d2-action-1}
\begin{split}
  \mathfrak{D}_1 \: = \: - \:
  Tr \Big(
  & [A,B] [\Delta^A,\Delta^B] \: + \:
  [B,A] [\Delta^B,\Delta^A] \: + \: \\
  & [A,B] [\Delta^B,\Delta^A] \: + \:
  [B,A] [\Delta^A,\Delta^B] \: + \: \ldots
  \Big)
\end{split}
\ee
and after expanding the trace $Tr$ this gives rise to
\be \label{d2-action-2}
  \mathfrak{D}_1 \: = \:
  -2 \: tr \left(
  B_{i+3,j+2} A_{i+2,j+1} \Delta^B_{i+3,j+3} \Delta^A_{i+3,j+2}
  \: - \:
  B_{i+1,j+1} C_{i,j} \Delta^C_{i,j} \Delta^B_{i+1,j+1}
  \: + \: \ldots
  \right)
\ee%
 the trace is now over $U(N)$ generators. The factor of $2$
results from the equivalence of $[A,B] [\Delta^A,\Delta^B]$ and
$[B,A] [\Delta^B,\Delta^A]$, both of which appear in the sum in
$\mathfrak{D}_1$. The relative sign between the two terms
displayed in \eqref{d2-action-2} reflects the fact that they
arise, respectively, from the first and second line of
\eqref{d2-action-1}, which differ in their relative ordering of
fields in the commutators.

To see how interaction plaquettes make their appearance, consider the operators
\be
  O_1 \: = \: tr
  \left(
  \: A_{i+3,j} \: A_{i+3,j+1} \: A_{i+3,j+2}
  \: B_{i+3,j+3} \: B_{i+2,j+2} \: B_{i+1,j+1}
  \: C_{i,j} \: C_{i+1,j} \: C_{i+2,j}
  \right)
\ee
and\footnote{These are taken to be local operators in the
$3+1$ dimensional space-time, so the fields all sit at the same
space-time point.}
\be
  O_2 \: = \: tr
  \left(
  \: A_{i+3,j} \: A_{i+3,j+1} \: B_{i+3,j+2}
  \: A_{i+2,j+1} \: B_{i+2,j+2} \: B_{i+1,j+1}
  \: C_{i,j} \: C_{i+1,j} \: C_{i+2,j}
  \right)
\ee
for some fixed $i,j$,
and study the action of $\mathfrak{D}_1$ on $O_1$. We have
\be \label{d2-action-3}
  \mathfrak{D}_1 \: O_1 \: = \:
  2 \: N
  \left(
  3 O_1 \: - \: O_2 \: + \: \cdots
  \right)
\ee
with $\cdots$ representing other operators we are ignoring for now.
Equation \eqref{d2-action-3} also shows that a factor of $N$ arises from
merging the traces in $\mathfrak{D}_{1}$ and $O_1$ ($O_1$ is a single trace operator),
and the $3$ counts the number of corners in $O_1$, since the action of the
dilatation operator is localized only at corners and straight parts of operators are
not acted on by the
dilatation operator. The factor of $N$ combines with the $g^2_{YM}$ in
\eqref{extract-coupling} to give the effective 't Hooft coupling. Note that
$\mathfrak{D}_{1}$ is one-loop {\it planar} dilatation operator.

\begin{figure}[ht]
\centering
\epsfig{figure=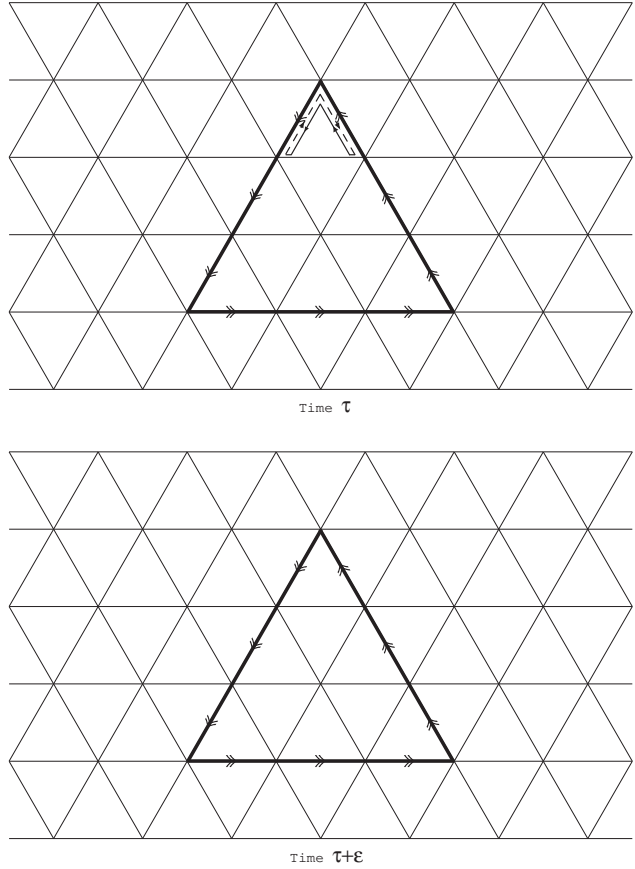,width=236.25pt,height=329.73pt}
\begin{center}
\caption{The action of a corner plaquette term at a corner in a single time step.
This contributes +2 to the transition amplitude.}
\label{interact-3}
\end{center}
\end{figure}

\begin{figure}[ht]
\centering
\epsfig{figure=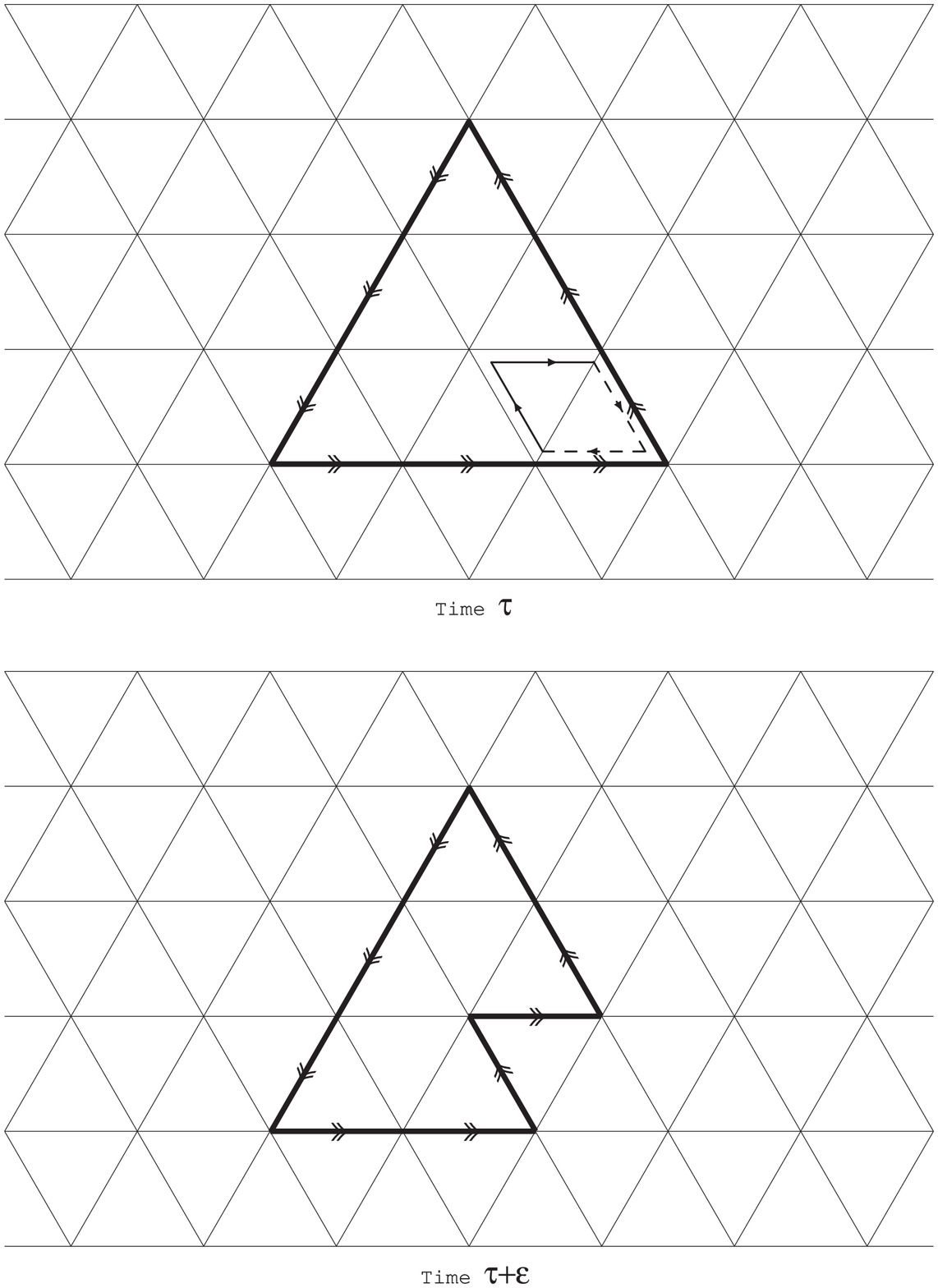,width=236.25pt,height=329.73pt}
\begin{center}
\caption{The action of an open plaquette term at a corner in a single
time step. This contributes -2 in units of $g^2_{YM}N$ to the
transition amplitude.}
\label{interact-1}
\end{center}
\end{figure}

Figures \ref{interact-1} and \ref{interact-3} depict equation
\eqref{d2-action-3} graphically.
The general structure of interactions is as follows: each term in the expansion of $\mathfrak{D}_1$
can be represented diagrammatically as a plaquette. Such plaquettes can be inserted anywhere
where the two dashed lines, which correspond to the derivatives in $\mathfrak{D}_1$, can be
contracted with fields in an operator (the arrows on the fields and the derivatives with
which they contract must run in opposite directions).\footnote{The directions of the arrows
are due to the fact that $\Delta^I$ has the
$U(N)$ transformations of $\bar{\Phi}^I$, as they arise from Wick contractions in Feynman diagrams where the field $\Phi^{I}$ is contracted with $\bar{\Phi}^{I}$, and
the derivatives in $\mathfrak{D}_1$ are a shorthand way of capturing this.}
The contracted fields disappear in the next time increment, to be replaced by the two
remaining fields in the plaquette. The amplitude for such a transition is a numerical
coefficient associated with the plaquette, and this defines the matrix element of the
transfer matrix between these states. Of course there are in general many terms in the
expansion of $\mathfrak{D}_1$ which have non-vanishing action when acting on a generic operator.
Each such term gives rise to a possible transition, with the associated amplitude.

In general, every holomorphic or anti-holomorphic plaquette
does one of two things: it either commutes two fields in the operator
(adjusting the lattice indices appropriately), or leaves the order of
the fields unchanged. This is obvious from the structure of the plaquettes.

The earlier argument that holomorphic operators form a closed
subset can now be seen graphically. The only plaquettes which can
contract into such operators must have two holomorphic derivatives
(i.e.\! derivatives with respect to holomorphic fields), and as
pointed out in Appendix A, these derivatives transform in the
conjugate representation, for which the arrows run in the opposite
direction. The two fields in the plaquette must then be
holomorphic. As a result, the insertion of such a plaquette
absorbs two holomorphic fields and replaces them with two
holomorphic fields, and hence holomorphic operators transition to
holomorphic operators. These considerations apply in an obvious
fashion to anti-holomorphic operators as well. Plaquettes
containing both holomorphic and anti-holomorphic fields also
contain both holomorphic and anti-holomorphic derivatives, and
their contraction with purely holomorphic or anti-holomorphic
operators vanishes. These considerations also imply that mixed
operators can never evolve to holomorphic or anti-holomorphic
operators.

Recall also our definition of BPS operators. These are operators that are constructed
solely from one of the three holomorphic (likewise anti-holomorphic) fields alone, and
wrap the lattice any number of times more than zero, in a gauge-invariant
way (they
start and end at the same lattice site).
The only plaquettes which could in principle be contracted with these operators must have
two derivatives with respect to the same field, both holomorphic or anti-holomorphic, but not mixed. A glance back at equation \eqref{n4-dilatation} shows immediately that no such
plaquettes exist, as $[\Delta^A,\Delta^A]$, etc. vanishes identically. This justifies the
term BPS we introduced earlier.

The evolution generated by the transfer matrix is (imaginary) time reversal
symmetric, in the following sense:
for any evolution from a string configuration associated to operator $O_1$ at time $\tau$
to configuration $O_2$ at time $\tau+\epsilon$ generated by a plaquette $p$, there exists
a plaquette $\bar{p}$
that would transform configuration $O_2$ at time $\tau$ to $O_1$ at
time $\tau+\epsilon$, or in other words, if we run time backwards,
the string configuration $O_2$ at time $\tau+\epsilon$ is taken to
configuration $O_1$ at time $\tau$ by the plaquette $\bar{p}$.
The plaquette $\bar{p}$ which accomplishes this is gotten by flipping the
plaquette $p$ along the diagonal as depicted in
Figure \ref{transform1} for zero area plaquettes,
while for the closed plaquettes $p=\bar{p}$,
as their action is proportional to the
identity operator.

\begin{figure}[tbph]
\centering
\epsfig{figure=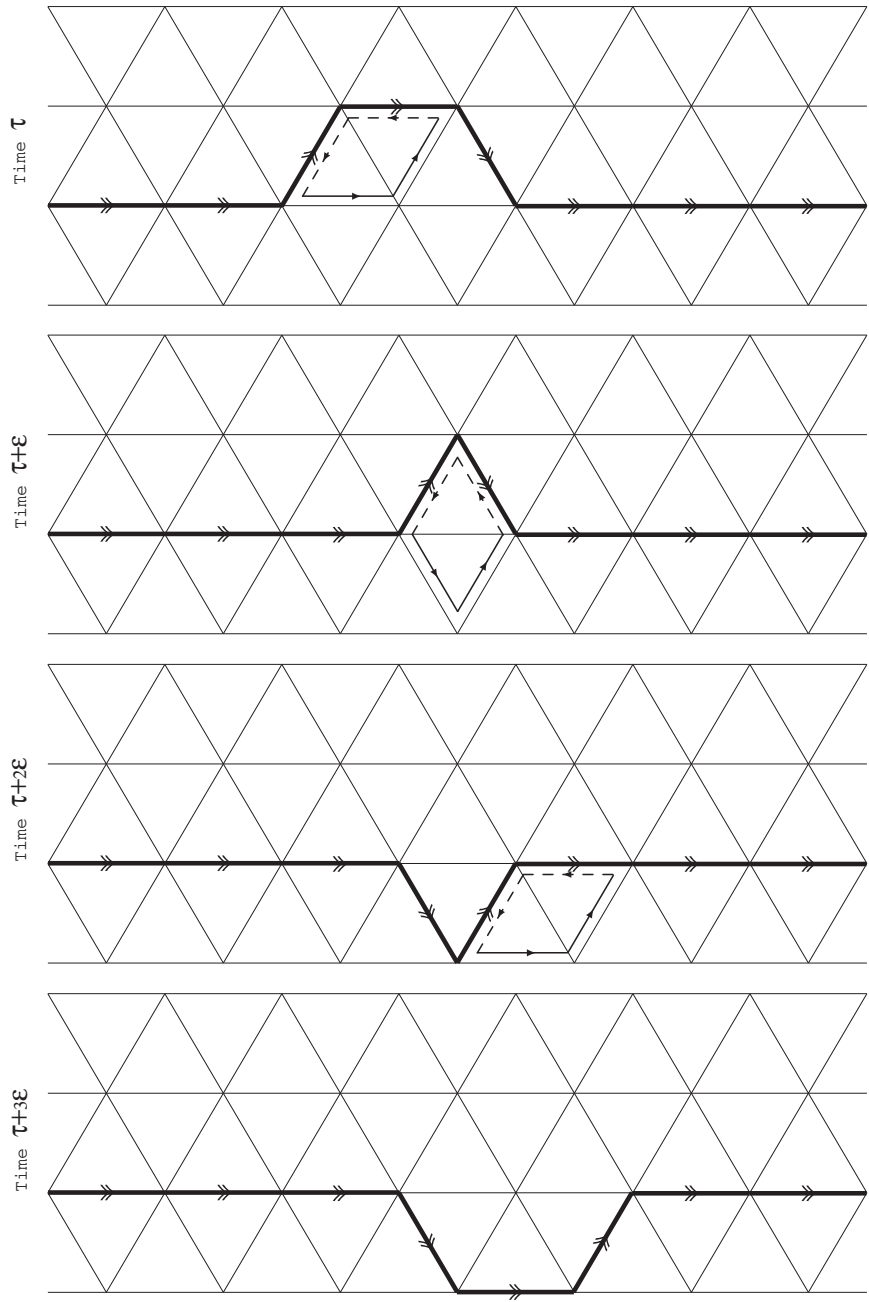,width=327.56pt,height=490pt}
\begin{center}
\caption{Following some possible string fluctuations for several time
steps. This describes a wave propagating along the string.}
\label{interact-2}
\end{center}
\end{figure}

A more elaborate example of wave propagation is shown in
Figure \ref{interact-2}, with the configuration of the string depicted at
four instants of time, and the
plaquettes generating the transitions also displayed.
Note that some of the plaquettes which appear in this example mix
holomorphic and anti-holomorphic fields.

It should be clear from this example that (planar level)
fluctuations of strings can only take place at corners. In other
words, the straight portions of strings can not be deformed. This
behavior is clearly the reason BPS operators are protected.

The structure of the interaction plaquettes also make it evident that the string
length is unchanged when it fluctuates. This is simply a restatement of the observation
that the one-loop dilatation operator $\mathfrak{D}_1$ only connects operators of the same dimension,
since for operators built only from the scalar
bi-fundamental fields the dimension is the same as the length. (This generalizes
in an obvious way to the fermion bi-fundamentals, if we take account of their canonical
dimensions appropriately.)
For the case of holomorphic or anti-holomorphic operators, this is also the statement
of $R$-charge conservation, as noted previously.
\begin{figure}[ht]
\centering
\epsfig{figure=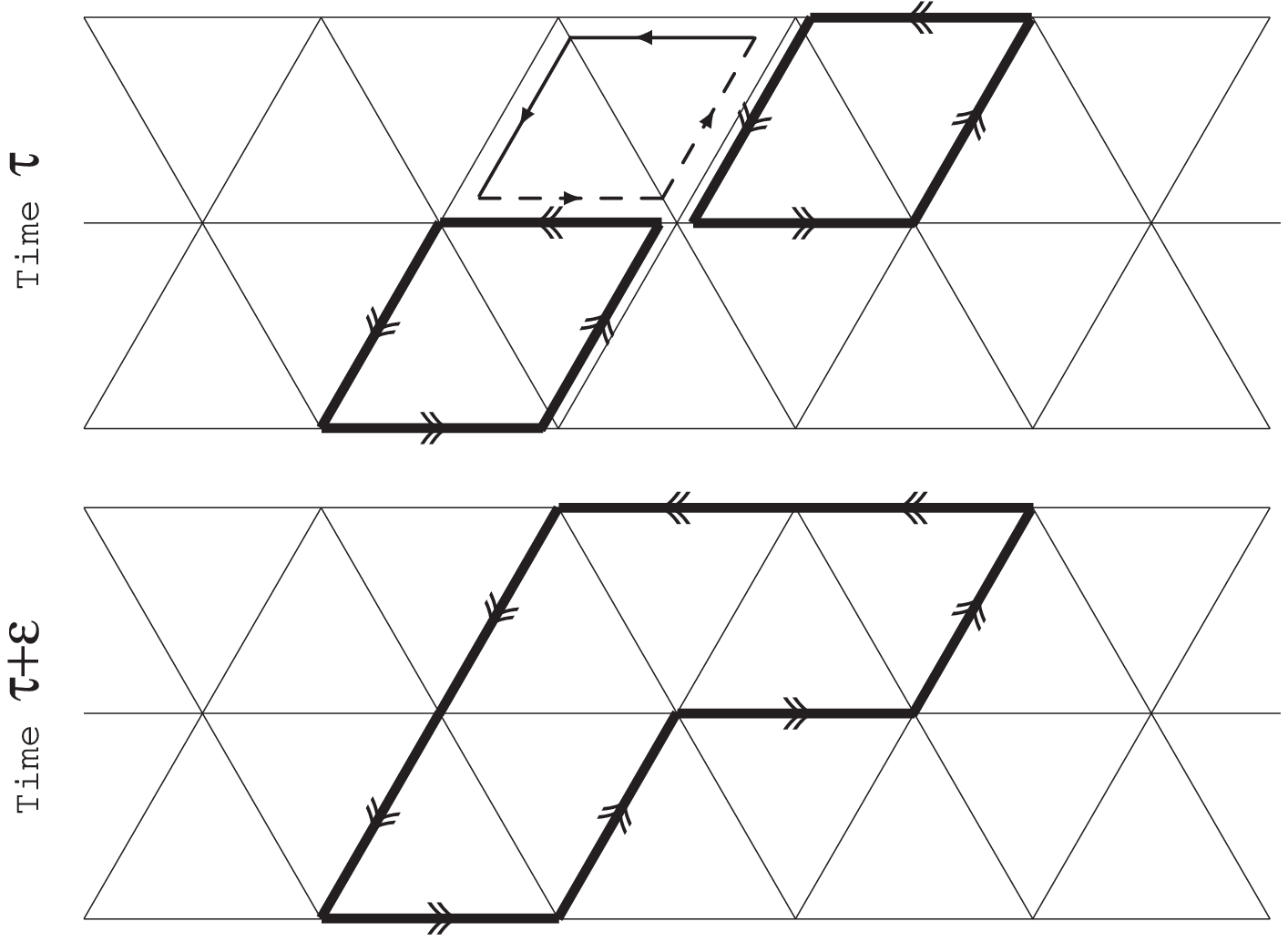, width=277pt,height=200pt}
\begin{center}
\caption{An example of a $1/N$ string interaction diagram,
where a double trace operator merges with a single trace one.
The $1/N$ suppression is appropriate to (bi)-fundamental fields.}
\label{join}
\end{center}
\end{figure}

\section{Dynamical Pictures}
\label{dynamical-pictures}

In the previous section we discussed that how the gauge invariant
operators of the $\cN=1$, and in particular the holomorphic operators,
can be realized on the two dimensional Moose diagram. We also outlined
how to perform one-loop planar computations of two point functions, and hence
anomalous dimension matrix, in a pictorial way, using the overlaps of operators
on the lattice. In this section we  build a closer connection with
the spin chain on the Moose and/or the lattice theory picture.

\subsection{Lattice Laplacian Picture}
\label{lattice-laplacian-picture}

In section \ref{transfer-matrix} we drew an analogy between the one-loop dilatation operator and the transfer matrix
describing (Euclidean) time evolution.
Wave propagation on a string is governed by a wave equation. We demonstrate here
that a basis of operators can be chosen such that their time evolution is described
by a Laplacian, together with extra contact terms,
giving another perspective on the dynamics of the loops, and through our
dictionary, the anomalous dimension matrix.

\begin{figure}[th]
\centering
\epsfig{figure=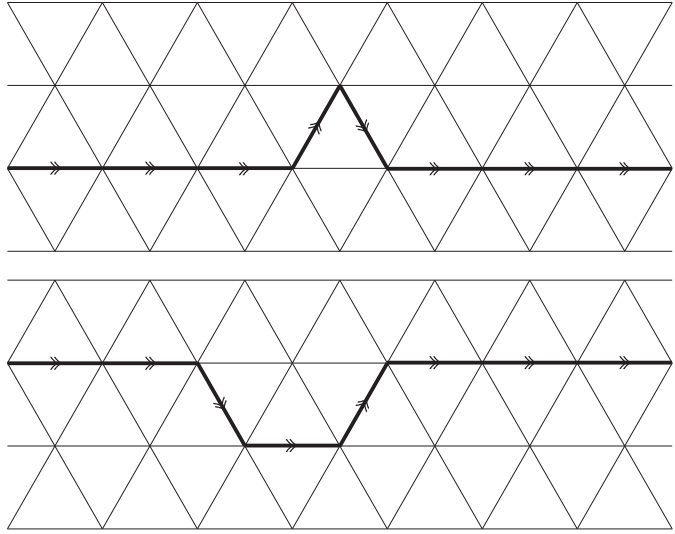,width=252pt,height=200pt}
\begin{center}
\caption{Examples of operators with winding number one, on a
lattice of size $l=7$ in the $A$ direction, where we assume
periodic boundary conditions at the edges.
Both operators have length or dimension $d=l+1=8$.
In the first case, $i=4,j=5$, so we denote it as
$O_{4,5}$. In the second example $i=5,j=3$, so this operator is
labeled $O_{5,3}$.}
\label{laplace1}
\end{center}
\end{figure}

In this subsection we give a description of the string evolution in terms of solutions of a
Laplace equation. First we must introduce a basis of operators.
For concreteness, we focus on a subsector of operators $O_{i,j}$ of
the form
\be \label{O-def}
  O^l_{i,j} \: = \:
  tr \big( \: \overbrace{\underbrace{\overbrace{
  A \ \ \ldots \ \ A}^{i-1} \ \bar{C} \ A \ \ \ldots \ \ A}_{j-1}
  \ \bar{B} \ A \ \ \ldots \ \ A}^{p+2} \: \big)
\ee with the operator $\bar{C}$ located in the $i$'th position,
and the operator $\bar{B}$ located in the $j$'th. The superscript
$l,\ l=1,2,\cdots, q$, denotes where (most of) the A's line lies
in the $q$ direction of the $2d$ lattice. In principle, the
operators of the form \eqref{O-def} can have a ``winding number'',
counting the number of times the operator wraps the lattice. We
focus on operators of winding number one, the generalization to
higher winding modes goes through with the most obvious
modifications. In $O^l_{ij}$ there are $i-1$ $A$'s appearing
before $\bar{C}$ and $j-1$ before $\bar{B}$, but we do not assume
that $\bar{C}$ appears before $\bar{B}$, so that both cases where
$i<j$ and $i>j$ are allowed (but not $i=j$). The total length of
operators of this form with winding number one is $p+1$, with $p$
the size of the lattice in the $A$ direction. Examples of such
operators are depicted in Figure \ref{laplace1}.

Recall that operators built strictly from $A$'s would be BPS, and would receive no
anomalous dimension corrections. The presence of $\bar{B}$ and $\bar{C}$ causes this operator
to no longer be BPS, but since fluctuations are allowed only to occur at corners and not
along the straight portions of operators, this operator is in some sense
almost-BPS in the large $p$ limit, as
the portions of the operator away from the insertions of $\bar{B}$ and $\bar{C}$ remains
BPS. These insertions are the $\mathcal{N}=1$ analogues of
BMN-type impurities in the $\mathcal{N}=4$ theory.

An example of the evolution of such operators has already been
shown in Figure \ref{interact-2}, where the transitions are
$O^l_{3,5} \rightarrow O^l_{4,5} \rightarrow O^l_{5,4} \rightarrow
O^l_{6,4}$. As can readily be seen from Figure \ref{interact-2}
and is inferred from discussions of previous section the
dilatation operator $\mathfrak{D}_1$ does not act on the $l$ index
and hence hereafter we drop the $l$ index and simply denote the
operators of the form \eqref{O-def} by $O_{i,j}$.

 For a general operator $O_{i,j}$ of the form \eqref{O-def}
the action of the dilatation operator is given by\footnote{
Equation \eqref{dilatation-op-action} applies for any winding
number for two-impurity operators, and so we drop the winding
number when writing the operator. The tree-level dilatation
operator does care about the winding number however.} \be
\label{dilatation-op-action}
\begin{aligned}
  \mathfrak{D}_1 \: O_{i,j} \: = \:
  &+4 \: O_{i,j} \: - \:
  O_{i-1,j} \: - \:
  O_{i+1,j} \: - \:
  O_{i,j-1} \: - \:
  O_{i,j+1} \\
  &+ \delta_{i+1,j} \left( O_{i,j-1}+ O_{i+1,j} - O_{i,j} - O_{i+1,j-1} \right) \\
  &+ \delta_{i-1,j} \left( O_{i,j+1}+O_{i-1,j}  - O_{i,j} - O_{i-1,j+1} \right)
\end{aligned}
\ee
with the constant $4$ on right-hand side counting the number of corners in the operator.\footnote{In general,
such operators will have four corners, except when the impurities $\bar{B}$ and $\bar{C}$
appear next to each other, in which case $j=i \pm 1$. This later case has three corners, and the contact term corrects for this.}
The second line is a contact term which takes account of the configuration where the two
impurities sit next to each other. The $O_{i,j}$ in the contact term correct the number of
corners and the last term accounts for the flip transitions when a bump pointing up
transitions to a bump pointing down and vice-versa.

We would now like to show the appearance of a latticized Laplacian. To facilitate the
rewriting, we introduce the forward and backward shift operators acting the first or
second index
\be
\begin{split}
  \mathbb{S}_i \: O_{i,j} \: = \: O_{i+1,j} \ ,
  \ \ \ \ \ \
  \ \ \ \ \ \
  \hat{\mathbb{S}}_i \: O_{i,j} \: = \: O_{i-1,j} \ , \\
  \mathbb{S}_j \: O_{i,j} \: = \: O_{i,j+1} \ ,
  \ \ \ \ \ \
  \ \ \ \ \ \
  \hat{\mathbb{S}}_j \: O_{i,j} \: = \: O_{i,j-1} \ ,
\end{split}
\ee
and the identity operator
\be
  \mathds{1} \: O_{i,j} \: = \: O_{i,j} \ .
\ee
In terms of these we define the forward and backward difference operators
\be
\begin{split}
  \nabla_i \:  \: \equiv \: \mathbb{S}_i \: - \: \mathds{1} \\
  \hat{\nabla}_i \:  \: \equiv \: \mathds{1} \: - \: \hat{\mathbb{S}}_i \\
  \nabla_j \:  \: \equiv \: \mathbb{S}_j \: - \: \mathds{1} \\
  \hat{\nabla}_j \:  \: \equiv \: \mathds{1} \: - \: \hat{\mathbb{S}}_j
\end{split}
\ee
from which we also define two Laplacian operators acting on $i,j$
\be
\begin{split}
  \nabla^2_i \: \equiv \: \nabla_i \: - \: \hat{\nabla}_i \\
  \nabla^2_j \: \equiv \: \nabla_j \: - \: \hat{\nabla}_j
\end{split}
\ee
and the total Laplacian
\be
  \nabla^2_{i,j} \: \equiv \: \nabla^2_i \: + \: \nabla^2_j
\ee

With these definitions, we can rewrite the dilatation operator,
when acting on the almost-BPS operators
\eqref{dilatation-op-action} as \footnote{Note that the actual
one-loop anomalous dimension is given by $g_{YM}^2 N/(4 \pi)^2
\mathfrak{D}_1$ after restoring the coupling we extracted in
\eqref{extract-coupling}.} \be \label{dilatation-laplace}
  \mathfrak{D}_1 \: O_{i,j} \: = \:
  \Big( \nabla^2_{i,j} \: - \:
  \delta_{i+1,j} \nabla_i \: \hat{\nabla}_j \: - \:
  \delta_{i-1,j} \hat{\nabla}_i \: \nabla_j
  \Big) O_{i,j}
\ee%
The second and third term above are contact terms which
correct for the case when we have a minimal size bump, in which
case a bump pointing up can flip to a bump pointing down and
vice-versa, and also the fact that for minimal size bumps there are
only three corners instead of four.

In writing the action of the dilatation operator in this form, we
have made evident the picture of operator evolution in terms of
waves propagating on a fluctuating string manifest.


To find the eigenvalues and eigenstates of $\mathfrak{D}_1$ let us
focus on \eqref{dilatation-op-action}. The cyclicity of the trace
in \eqref{O-def} is the remnant of the translational
invariance of the lattice in the $A$ direction. This implies that
the eigenstates of \eqref{dilatation-op-action} should only be a
function of $i-j$. Defining $k\equiv i-j$ (note that the $k$ index
does not take the value 0) then
\eqref{dilatation-op-action} takes the form%
\be\label{dil-opt-k}
\begin{split}
\mathfrak{D}_1 O_{k}&=4O_k -2(O_{k-1}+O_{k+1})\ ,\ \ \ \ k\neq\pm 1\\
\mathfrak{D}_1 O_{+1}&=3O_{+1} -2O_{+2}-O_{-1}\\
\mathfrak{D}_1 O_{-1}&=3O_{-1} -2O_{-2}-O_{+1}
\end{split}
\ee%
Next, let us define%
\be\label{Opm}%
O^{\pm}_k= O_k\pm O_{-k} \ .%
\ee%
In terms of $O^\pm_k$ \eqref{dil-opt-k} the equations for $O^+$ and
$O^-$ decouple:%
\begin{subequations}\label{dil-opt-k-pm}
\begin{align}
\mathfrak{D}_1 O^{\pm}_{k}&=4O^{\pm}_k -2(O^{\pm}_{k-1}+O^{\pm}_{k+1})\ , \ \ \ k\neq\pm 1 \ ,\\
\mathfrak{D}_1 O^+_{1}&=2(O^+_{1} -O^+_{2}) \ ,\\
\mathfrak{D}_1 O^-_{1}&=4O^-_{1} -2O^-_{2} \ . %
\end{align}
\end{subequations}
If we define %
\be\label{boundary-conditions}%
O^+_0\equiv O^+_1\  ,\ \ O^-_0\equiv 0 \ ,
\ee%
then equations (\ref{dil-opt-k-pm}b,c) become compatible with
(\ref{dil-opt-k-pm}a) once we relax the $k\neq \pm 1$ condition
and allow $k$ to also take the value 0. In fact
\eqref{boundary-conditions} plays the role of ``boundary
conditions'' for the Laplace equation of motion for $O^{\pm}_k$'s,
$k\geq 0$; $O^+_k$ have a Neumann boundary condition and $O^-_k$
Dirichlet. We should stress that  this ``boundary condition'' is
not related to the boundary conditions for the closed strings,
which still have periodic boundary conditions. Therefore,
\begin{subequations}\label{Qn}
\begin{align}
Q^+_n & \: = \: \sum_{k=1}^{p+1} O^+_k \cos(\frac{2\pi\ nk}{p+1}) \ , \\ %
Q^-_n & \: = \: \sum_{k=1}^{p+1} O^-_k \sin(\frac{2\pi\ nk}{p+1}) \ ,
\end{align}
\end{subequations}
and both have $\mathfrak{D}_1$ eigenvalues%
\be\label{eigenvalue}%
\omega^\pm_n \: = \:
\frac{g^2_{YM}N}{(2\pi)^2}\left(1-\cos(\frac{2\pi n}{p+1})\right) \ ,%
\ee%
where we have reintroduced the 't Hooft coupling
$\lambda=g^2_{YM}N/(4\pi)^2$.

\begin{figure}[th]
\centering
\epsfig{figure=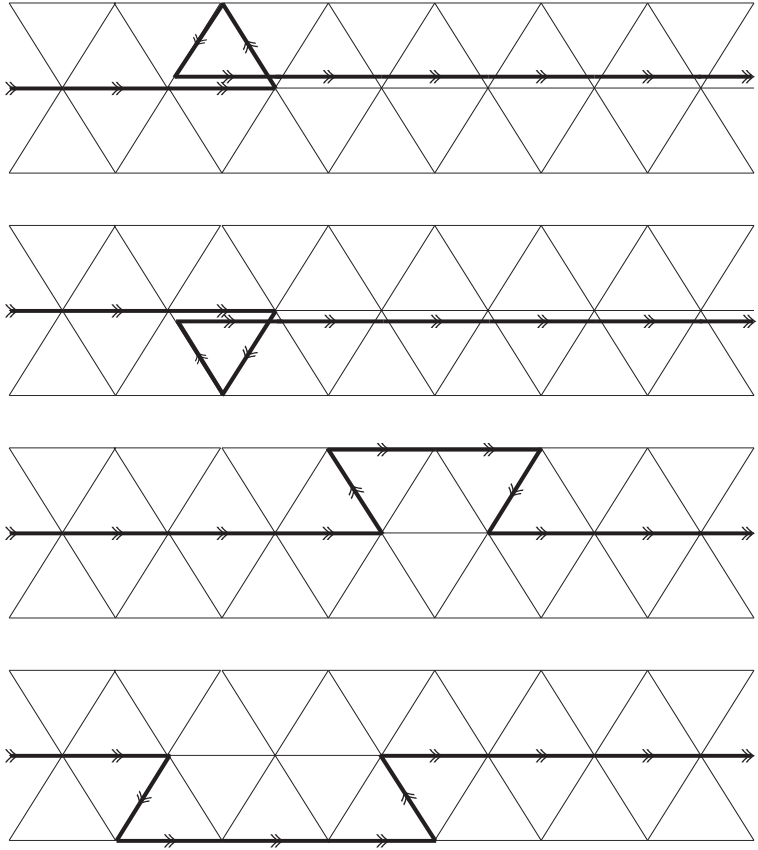,width=285pt,height=320pt}
\begin{center}
\caption{Examples of operators with winding number one, on a
lattice of size $l=7$ in the $A$ direction, where we assume
periodic boundary conditions at the edges. Both operators have
length or dimension $d=l+1=8$. In the first case, $i=4,j=5$, so we
denote it as $O_{4,5}$. In the second example $i=5,j=3$, so this
operator is labeled $O_{5,3}$.} \label{BMN-ops}
\end{center}
\end{figure}

One could repeat a similar analysis with operators of the form
\be
  \hat{O}^k_{i,j} \: = \:
  tr \left( B_{i,j} A^k C A^{p-k+1} \right) \ ,
\ee%
which are in the holomorphic sector. For these operators we again
find a result similar to the previous non-holomorphic two
impurity case. An example of such operators and their time
evolution is depicted in Figure \ref{BMN-ops}.

\subsection{The BMN Limit}
\label{BMN-limit}

It is instructive to consider the ``BMN'' limit of our $\cN=1$
theory and its realization on the two dimensional triangle
lattice. As this point has to some extent been discussed in the
literature, especially on the gravity side and when taking the Penrose
limit, we will be very brief and only address some issues related to the
the gauge theory. For some earlier works on the Penrose limit of
orbifolds, see
\cite{Semenoff,Compactified-ppwaves,Orbifolds-Penrose-limit,
worldsheet-deconst.}.

Let us consider the $p=q$ case. The BMN-type limit is then taking
$p, N\to\infty$ while keeping $g^2_{YM}$ and $p^2/N$ fixed.  This
BMN-type limit can be thought of as the continuum limit over the
discretized string worldsheet. Restricting to operators made out
of the bi-fundamental scalars, this is in fact also the continuum
limit over a 2+1 dimensional target space. The BMN-type
operators are those with dimension (or $R$-charge) of order of
$p$. The BMN vacuum states are the straight lines wrapping the
lattice. These operators may be in the $A$, $B$ or $C$ directions
and in the twisted or untwisted sector, e.g.
$tr(A_{i,j+1}A_{i,j+2}\cdots A_{i,j+p})\ ,\ i=1,2,\cdots, p$. As
discussed earlier, these operators are BPS and have vanishing
anomalous dimension and are labeled by two quantum numbers: their
dimension (or $R$-charge) $p$ and the $i$ index (or its Fourier
transform). Moreover, there is another possibility that this
operator wraps the lattice in the $A$ direction some number of
times. This ``winding'' $w$ is hence the third quantum number
needed to specify the vacuum state. (The dimension of the operator
is then $pw$.)

From the dual string theory viewpoint, this operator is the vacuum
state for the discrete light-cone quantization (DLCQ)
of string theory on the corresponding
plane-wave \cite{Compactified-ppwaves}. The string can only move
freely in one direction (other than the light-cone direction) and
is confined in the other directions due to the harmonic oscillator
potential coming from the background plane-wave. Therefore, the
vacuum state is only labeled by three quantum numbers, $p^+$, the
light-cone winding $w$, and a momentum which is the Fourier transform
of the $i$ index.\footnote{The fact that straight line operators,
e.g. of the form of $tr(A_{i,j+1}A_{i,j+2}\cdots A_{i,j+p})$ for
different $i$, have zero overlap is not strictly correct. To be
precise there is an overlap between the $i$ and $i+1$ operator at
$p$-loop level. This is not in conflict with momentum conservation
because the triangle lattice is sitting on a torus
and the lattice directions are compact hence the momentum along
them can have a jump by integer multiples of $p$. In the lattice
(field) theories this phenomenon is the well known {\it Umklapp}
effect. For large $p$ (i.e. decompactified torus), however, this does not
happen.}

The string excitations above the vacuum are then given by placing
small bumps on this straight line of $A$'s (as depicted in Figure
\ref{interact-2}). In the large $p$ limit these are almost BPS
BMN-type operators.

One may read off the effective 't Hooft coupling  for this case
from \eqref{eigenvalue}. The result of computations, when $p=q$,
are%
\be\label{effective-tHooft}
\lambda' =\frac{g^2_{YM} N}{p^2}\
\ee
where $\lambda'$ is the effective dressed 't Hooft coupling.
%
This result may be argued for by noting that the $\cN=1$
theory we are considering is obtained as a $\mathbb{Z}_p\times
\mathbb{Z}_p$ orbifold of a $U(Np^2)$ $\cN=4$ gauge theory for
which the effective 't Hooft coupling in the BMN sector with
$R$-charge $J$ is $\lambda'=
\frac{(g_{YM}^{parent})^2(Np^2)}{J^2}$. %
In our case $J=p$ and $g^2_{YM}$ and $(g_{YM}^{parent})^2$ are
related as in \eqref{N=1-coupling}. Note that this argument is
only applicable to the untwisted sector. However, as in the usual
string theory, one expects this to be valid for the twisted sector
as well. This expectation is indeed confirmed by explicit
computations we have shown in section
\ref{lattice-laplacian-picture}.

\subsection{Spin Chain Picture and Integrability}
\label{spin-chain}

In (Lorentzian) real space, the path integral gives the amplitude for a
configuration of the system at some
initial time to evolve to the prescribed configuration at the final time.  It involves summing over all allowed
intermediate states with a given weight, which is the exponential of minus $i$ times the action.
When going over to Euclidean space, $i$ times the action becomes minus the Euclidean action, which is just the energy.
So in Euclidean space the path integral for a $d+1$ dimensional system is really the partition function of a
classical $d+1$ dimensional statistical mechanics system. Using the transfer matrix, this can be related to
a quantum mechanical system in $d$ dimensions.  This involves taking a limit of the transfer matrix formulation
in which the time variable becomes continuous, tuning the spatial and temporal couplings in a special way \cite{Kogut}.

We have a Euclidean system in which a configuration of strings on a two-dimensional lattice evolves to another
configuration on the same two-dimensional lattice in a discrete Euclidean time step. We have already described how to compute the amplitude for such a transition, which is given by the matrix elements of the transfer matrix between the appropriate initial and final configurations.
As we discussed previously, the dimensionality of the transfer matrix is determined by the number of allowed possible string configurations on the lattice.
We would now like to relate this to a lower dimensional quantum mechanical system.
As usual, in the limit of infinitesimal time steps, the transfer matrix can be expanded as
\be \label{transfer-matrix-H}
  \hat{T} \: = \: 1 \: - \: \epsilon \hat{H}
\ee with $\epsilon$ the infinitesimal time step. By analogy to the
relationship between the classical two-dimensional Ising model and
the quantum one-dimensional Ising chain, we expect the quantum
Hamiltonian to describe a two-dimensional system. Such a
description will be briefly discussed in section \ref{LGT}, where
we will see that the description can be given in terms of a
2+1-dimensional Euclideanized Lattice gauge theory in a temporal
gauge, in the conformal fixed point.

However, because the natural description we have been using
for the state of the system at a given time is in terms of string
configurations, instead of the configuration of all link variables
on a slice, the lower dimensional system will naturally turn out to
describe  $(1+1)$-dimensional objects.

Again, as is the usual structure, in the infinitesimal time limit,
only transitions which involve zero or one ``flip'' contribute at first order in $\epsilon$.
The flips occur when a single insertion of the plaquettes in Figure \ref{open-plaquettes}
flip two legs of a triangle along a diagonal. Zero flips are due to the insertions
of plaquette terms in
Figure \ref{corners}.
Earlier we identified the matrix elements of the transfer matrix for a transition between initial and final states with
the matrix elements of the one-loop dilatation operator. If we identify $\epsilon$ with the coupling (up to a numerical factor arising from the commutator structure of \eqref{n4-dilatation})
$2 \lambda/(4 \pi)^{2}$, where $\lambda=g_{YM}^{2} N$ is the 't Hooft coupling, then the fact that only zero or single flip transitions contribute is a consequence of
keeping only one-loop contributions to the dilatation operator, i.e. \! weak 't Hooft coupling.
So we make this identification, together with a slight modification of our earlier construction of the transfer matrix, where we now identify the transfer matrix with the complete dilatation operator to one-loop, including both the tree level
and the one-loop contributions \eqref{extract-coupling}.
Inclusion of the tree-level classical dimensions allows us to extract the
identity term in \eqref{transfer-matrix-H}.

A string loop is formed from a series of links, labeled $L_{i}$,
for $i=1,\ldots,d$, with $d$ the total length of the loop. The
state of each link is described by a basis vector in a vector
space $\mathds{V}$. For a general operator, $\mathds{V}$ is a
six-dimensional  vector space  with basis vectors corresponding to
the fields $A,B,C,\bar{A},\bar{B},\bar{C}$ and hence the Hilbert
space of a loop of length  $d$ is $6^d$ dimensional. The basis for
$\mathds{V}$ can be decomposed into $\bf{3}\oplus \bar{\bf{3}}$ of
$SU(3)$. This $SU(3)$ is the subgroup of the $SU(4)_R$ of the
${\cal N}=4$ parent gauge theory before the orbifolding.

When restricting to holomorphic operators, we work in a
three-dimensional complex subspace of this vector space which
transform in ${\bf 3}$ or ${\bar{\bf 3}}$ of $SU(3)$. In the
holomorphic sector $\mathds{V}$ reduces to $\mathbb{C}^3$.
 We denote the
basis vectors for the $i$'th link as $\hat{e}_{i}^{\alpha}$, where
$\alpha$ ranges over $A,B,C,\bar{A},\bar{B},\bar{C}$. We denote
the state of the $i$'th link as $\hat{s}_{i}$. We introduce for
later use the permutation operator \be \label{perm-op}
  \hat{P}_{i,i+1} \ \hat{s}_{i} \otimes \hat{s}_{i+1} \: \equiv \:
  \hat{s}_{i+1} \otimes \hat{s}_{i} \ ,
\ee
This operator acts on tensor products of vector spaces associated to neighboring links, or alternatively on pairs of nearest neighbor links, and acts on the states at the positions $i$ and $i+1$ by exchanging them.
The full Hilbert space $\mathcal{H}$ of the string is the tensor product over the Hilbert spaces of each individual link
\be
  \mathcal{H} \: = \: \bigotimes_{i=1}^d \mathcal{H}_i\ .
\ee

Identify the matrix element of the transition matrix between
states $O_{i}$ and $O_{j}$ as \be \label{t-matrix-element}
  T_{i,j} \: = \:
  \big< \: \mathcal{O}_i \: | \: \hat{T} \: | \: \mathcal{O}_j \: \big> \: = \:
  \big< \: \mathcal{O}_i \: | \left( \cD_0 \: + \: \frac{g_{YM}^2}{(4 \pi)^2}
  \cD_1 \right)
  | \: \mathcal{O}_j \: \big>\ .
\ee
Normalize operators so that
\be
  \big< \: \mathcal{O}_i \: | \: \mathcal{O}_j \: \big> \: = \:
  \frac{1}{d} \delta_{ij}\ ,
\ee%
with $d$ the classical dimension of the operator. Then,%
\be\label{D-normalization}
  \big< \: \mathcal{O}_i \: | \: \cD_0 \: | \: \mathcal{O}_j \: \big> \: =
  \: \delta_{i,j}\ .
\ee%
 With this normalization \eqref{t-matrix-element} and
\eqref{Tmatrix-D1} are equal. Since the \opt s with different
classical dimension have zero overlaps, in first and all higher
loops and even at non-planar level, $\mathcal{H}$ is the Hilbert
space for $T$ and the set of all the \opt s on the $2d$ lattice
can be classified into \opt s of given classical dimension.
Moreover, for the same reason \eqref{D-normalization} is an
appropriate normalization. If $i=j$, then $\cD_{1}$ equals twice
the number of corners in $O_{i}$ and if $i \ne j$, then $\cD_{1}$
equals zero if the two operators cannot be connected by a single
plaquette, and equals minus two if they can.

Hereafter we only consider the holomorphic sector. Noting the
planar behavior of the transfer matrix when acting on the strings
in the holomorphic sector, it can be written in the following form
\be
  \hat{T} \: = \: \sum_{i=1}^{d}
  \left(
  1 \: + \:
  \frac{2 g_{YM}^{2} N}{(4 \pi)^{2}}
  \sum_{\alpha,\beta} \sum_{\delta,\rho} \
  C_{\beta \rho}^{\alpha \delta} \
  T_{i+1}^{\dagger \alpha}
  T_{i+1}^{\beta}
  T_{i}^{\dagger \delta}
  T_{i}^{\rho}
  \right)\ .
\ee
The indices $\alpha,\beta,\gamma,\delta$ label the states and range over $A,B,C$ for the holomorphic subsector.
Here the $T_{i}^{\dagger \alpha}$ is a creation operator acting in the Hilbert space associated with the
$i$'th link, producing the state indexed by $\alpha$. Likewise, $T_{i}^{\beta}$ is an annihilation operator
for the state indexed by $\beta$.
This form follows from requiring that the transfer matrix as an operator generates the matrix elements
via \eqref{t-matrix-element}.
The matrix $C_{\beta \rho}^{\alpha \delta}$ has value
\be
  C_{\beta \rho}^{\alpha \delta} \: = \:
  \left(
  \delta_{\beta}^{\alpha} \delta_{\rho}^{\delta}
  - \delta_{\rho}^{\alpha} \delta_{\beta}^{\delta} \right) \ ,
\ee
and the following symmetry
\be
  C_{\beta \rho}^{\alpha \delta} \: = \:
  C_{\rho \beta}^{\delta \alpha} \ .
\ee%
Note that these are in fact the spin operators which will
appear again below.

To see  how the spin chain structure arises, it is useful to
consider an example. Take the set of operators \be
\begin{split}
  O_{1} = Tr(AABBCC) \ , \\
  O_{2} = Tr(ABABCC) \ , \\
  O_{3} = Tr(AABCBC) \ , \\
  O_{4} = Tr(CABBCA) \ ,
\end{split}
\ee
which form a closed set under renormalization at one-loop planar level.
Consider now some matrix elements of the transfer matrix \eqref{t-matrix-element}
between these states
\be
  T_{1,1} \: = \:
  1 \: + \: 2 \frac{g_{YM}^{2}N}{(4 \pi)^{2}} \ \frac{\mathscr{C}}{d}
\ee%
where $\mathscr{C}$ is the number of corners in the operator
$\mathcal{O}_1$ (three in this example). We also have %
\be
  T_{2,1} \: = \:
  - 2 \frac{g_{YM}^{2}N}{(4 \pi)^{2}} \frac{1}{d}
\ee
We expand the transfer matrix to first order, using \eqref{transfer-matrix-H}, to find the Hamiltonian
\be
  \big< \: \mathcal{O}_1 | \: \hat{T} \: | \mathcal{O}_1 \: \big> \: = \:
  1 \: - \: \epsilon \:
  \big< \: \mathcal{O}_1 | \: \hat{H} \: | \mathcal{O}_1 \: \big>
\ee
and
\be
  \big< \: \mathcal{O}_2 | \: \hat{T} \: | \mathcal{O}_1 \: \big> \: = \:
  - \: \epsilon \:
  \big< \: \mathcal{O}_2 | \: \hat{H} \: | \mathcal{O}_1 \: \big>
\ee
Setting $\epsilon=\frac{2 g_{YM}^{2} N}{(4 \pi)^{2}}$,
allows us to identify the Hamiltonian $\hat{H}$ as
\be \label{integrable-H}
  \hat{H} \: = \: \sum_{i=1}^{d} \: \left(1 \: - \: \hat{P}_{i,i+1} \right)
\ee
where $\hat{P}_{i,i+1}$ is the permutation operator, the identity is understood to act on the tensor product of two vector spaces,
and we have periodically identified the boundaries.
Note that the infinitesimal time limit corresponds to small 't Hooft coupling, where the perturbative
expansion of the dilatation operator \eqref{extract-coupling} is valid.
In making this identification, we have made use of the observation that the number of corners in a loop equals the classical dimension minus the number of straight pieces in the loop.\footnote{A straight piece is defined as any part of the loop where the state at the location $i$' matches that at location $i+1$.}
In general the number  of corners is not a conserved quantity, so the other operators with which
a given operator can mix may contain different numbers of corners, but it
is generically true (for holomorphic operators and at planar level) that the number of corners fixes the number of other operators with which mixing occurs.

The Hamiltonian \eqref{integrable-H} is the Hamiltonian of an
integrable (anti-ferromagnet) $SU(3)$ spin chain
\cite{Minahan-Zarembo,Wang:2003cu,Chen:2004mu}. The integrability
of the (anti-)ferromagnetic $SU(N)$ spin chain, construction of
the Lax pairs and transfer matrix and the infinite number of
commuting conserved charges in terms of the Lax pairs has been
done in \cite{KR81} and may also be found in the Appendix A of
\cite{Wang:2003cu}. Moreover, using the algebraic Bethe ansatz
equations eigenvalues of the Hamiltonian has been given in terms
of the rapidity parameter of the Bethe ansatz \cite{KR81}.
Therefore, here we do not repeat  the integrability arguments.

A nice representation of the permutation operator \eqref{perm-op}
and hence the spin chain Hamiltonian \eqref{integrable-H} can be
given as follows: since the state at each link is a vector in the
complex vector space $\mathds{V}$ of dimension six, we can
introduce, by analogy to the spinors, the spin operator acting on
$\mathds{V}$, with the representation %
\be \label{spin-operators}
  S^{ab}_{ij} \: = \: \delta^{a}_{i} \: \delta^{b}_{j} \: - \: \delta^{a}_{j} \delta^{b}_{i}
\ee%
with $(a,b)$ ranging from $1,\ldots,6$, and the indices
$(i,j)$ being the matrix indices for this six-dimensional
representation.
Restricting ourselves to holomorphic operators, we can project to
the complex three-dimensional subspace of $\mathds{V}$ spanned by
$A,B,C$. Taking a three-dimensional representation of the spin
operators \eqref{spin-operators}, we can rewrite the Hamiltonian
as \be\label{spin-spin-2}
  \hat{H} \: = \: 2 \: + \:
  \frac{1}{2} \left( S^{ab}_{l} S^{ab}_{l+1} \right) \: - \:
  \frac{1}{4} \left( S^{ab}_{l} S^{ab}_{l+1} \right)^2 \ .
\ee%
This is the Hamiltonian for an integrable $SU(3)$ spin chain
\cite{KR81,Faddeev96, Wang:2003cu}.\footnote{It is worth noting
that it is the dilatation \opt\ in the holomorphic sector which
corresponds to an $SU(3)$ integrable anti-ferromagnetic spin chain.
The full dilatation \opt\, however, is not
related to a known integrable system \cite{Wang:2003cu}.}
 It is important to point out that the counting we
have presented is particular to the case of holomorphic operators;
the additional complication with non-holomorphic operators can be
traced to the weights appearing in Figure \ref{non-hol}.

So far we have shown, basically by construction, that the
Hamiltonian of a spin chain system with certain
nearest neighbor interactions is equivalent to the one-loop planar
dilatation operator of the $\cN=1$ quiver gauge theory in the
sector with operators made only out of three kinds of
bi-fundamental scalars. As discussed, the equivalence is most
simply seen and established in the basis where we label our states
by the oriented closed loops on the two dimensional lattice.

One may try to rephrase the above statement in the language of the path
integral and partition functions of the two sides. The partition function
of the two dimensional statistical mechanical system, with $\cO_i$ and
$\cO_f$ as the initial and final states, is defined as
\be\label{spin-chain-partition}
\begin{split}
Z_{{\rm spin\ chain}}[\beta; \cO_i, \cO_f]&=\langle \cO_f| e^{\beta \hat H}
|\cO_i\rangle\cr
&=  \sum_{\{\mathcal{O}_{j_1}\} }\cdots \sum_{\{\mathcal{O}_{j_M}\} }
 \langle  \mathcal{O}_f  |
  \hat{T}
   |  \mathcal{O}_{j_1} \rangle
  \langle \mathcal{O}_{j_1}  |
  \cdots
  | \mathcal{O}_{j_M}  \rangle
   \langle \mathcal{O}_{j_M}  |
  \hat{T}
   |  \mathcal{O}_i \rangle\ ,
\end{split}
\ee where $\{\mathcal{O}_{j_k}\}$ denote the set of all closed
loops on the lattice. We should also take $M\to\infty$ at the
end of the computation. Each sum over $\{\mathcal{O}_{j_k}\}$ can
in turn be decomposed into sums over sets containing \opt s of
definite classical dimension.

On the other hand, in the $\cN=1$ SYM, if we restrict ourselves to insertions of
operators consisting only of scalars, then the transition amplitude between the
initial and final states $\cO_i$ and $\cO_f$ is
\be\label{SYM-partition}
\begin{split}
Z_{2d\ {\rm reduced\ SYM}}[g^2_{YM}N; \cO_i, \cO_f]&=
\int [D\Phi]\ e^{-S_{reduced}}\ \cO_i\ \cO_f\cr
&=\sum_{\{gauge\ inv.\ opt.\}}
\langle \cO_f| e^{-\cD}|\cO_i\rangle\ ,
\end{split}
\ee
where the subscript reduced stands for the reduction to a sector of scalars and
the gauge invariant operators in this sector are identified with the
orientable closed loops on the lattice, owing to the
block diagonal nature of the transfer matrix (or equivalently
$\mathfrak{D}_1^h$) described above.

It is interesting to see if the above correspondence between the
2+1 dimensional spin chain and the holomorphic sector (or more
generally the sector made out of scalars) of the $\cN=1$ SYM
theory can be extended beyond this sector to also include the gauge
fields. In  sections \ref{branebox} and
\ref{deconstruction}  we will argue that this may be
achieved if we view the 2+1 theory to as part of a
3+(2+1) six-dimensional theory.

\section{Relation to AdS/CFT and Brane Box Models}
\label{branebox}

As  discussed earlier, the $\cN=1$ $U(N)^{pq}$ gauge theory of interest can be obtained from an $\cN=4$ $U(Npq)$ SYM theory, via a specific
$\mathbb{Z}_p \times \mathbb{Z}_q$ orbifolding.
The action of this orbifold on the scalars appears as a non-trivial
$U(Npq)$ gauge rotation (which is not in a subgroup of $U(N)^{pq}$) while for
the gauge fields there is no such twist.

The above orbifolding can also be understood from  the gravity (string
theory) dual to the $U(Npq)$ $\cN=4$ SYM. The gravity dual
background in this case is $AdS_5\times S^5$ with
$R_{S^5}^4=l_s^4g_s Npq$. The orbifolding is then acting on the
$S^5$ part.  To see this, consider the $S^5$ embedded in a
$\mathbb{C}^3$ as $\sum_i z_i^2=R^2$. The action of $\mathbb{Z}_p
\times \mathbb{Z}_q$ on $z_i$ is%
\be\label{orbifolding}%
\mathbb{Z}_p:\left\{\begin{array}{ccc}
z_1\equiv e^{\frac{2\pi i}{p}} z_1\\ \ \ \ \ \\
z_2\equiv e^{-\frac{2\pi i}{p}} z_2\end{array}
\right.
\mathbb{Z}_q:\left\{\begin{array}{ccc}
z_1\equiv e^{\frac{2\pi i}{q}} z_1\\ \ \ \ \ \\
z_3\equiv e^{-\frac{2\pi i}{q}} z_3\end{array}
\right.
\ee%
The $AdS_5\times S^5/\mathbb{Z}_p \times \mathbb{Z}_q$ space can
also be obtained as the near horizon geometry of a stack of $N$
D3-branes probing a $\mathbb{C}^3/\mathbb{Z}_p \times
\mathbb{Z}_q$ singularity, where the branes are sitting at the $z_i=0$
fixed point. {}From this brane setup it is evident that the dual
theory should be a $U(N)^{pq}$ gauge theory with bi-fundamental matter fields,
as the $z_i=0$ is the point where the stack of branes and their
orbifold images are coincident. The brane setup also sheds light
on \eqref{N=1-coupling}, recalling the notion of fractional branes
\cite{fractional-brane} and the fact that the RR-charge (or
effective tension) of the branes at the orbifold is
$\frac{1}{l_s^4g_s}\cdot\frac{1}{pq}$ and that the tension, which
is the coefficient in front of the Born-Infeld action for the
brane, is (the inverse square of) the coupling for the low energy
Yang-Mills theory living on the brane. Moving the branes away from
the orbifold fixed point corresponds to moving to the Higgsed
phase (Coulomb branch) of the $U(N)^{pq}$ theory
\cite{fractional-brane} where the conformal symmetry is also
broken. We will come back to this point in the next section. One
may also try to take the Penrose limit(s) on the $AdS_5\times S^5$
geometries; this has been carried out for example in
\cite{Compactified-ppwaves}.

The two dimensional lattice is most easily seen in the T-dual
picture of the above orbifold scenario. Let us denote the angular
parts of $z_1/z_2$ and $z_1/z_3$ by $\alpha$ and $\beta$. These are
the two $S^1$ directions where the orbifolding acts. Now, perform
two T-dualities on the $\alpha$ and $\beta$ directions. The stack
of $N$ D3-branes is  mapped to $N$ D5-branes. The metric of the
$\mathbb{C}^3/\mathbb{Z}_p \times \mathbb{Z}_q$ has off-diagonal
pieces once two of the coordinates are chosen along the $\alpha$ and
$\beta$ directions. These off-diagonal terms upon T-duality become
NSNS $B$-fields whose three-form flux, in the near horizon
geometry we are interested in, corresponds to intersecting smeared
NS5-branes. The above can be summarized in the following
simplified five-brane setup: Consider a stack of $N$ D5-branes
along the 012345 directions and two sets of $p$ (and $q$) NS5-branes
along the 012346 (and 012357) directions. The NS5-branes are
respectively smeared in the 4 and 5  directions, and the D5-branes
are localized in the 6789 directions. The 45 plane is covered by
the $\alpha$ and $\beta$ directions mentioned earlier and is
wrapping a two torus. The $\alpha$ and $\beta$ directions do not
form an orthogonal basis for this torus.

The above intersecting brane setup leads to a generalization of
the Hanany-Witten type brane configuration \cite{HW}, where the
D5-branes now have a finite extent in two directions. This forms
the brane box picture \cite{Hanany-Uranga,Aharony:1997bh,
Hanany:1997tb,Erlich:1999rb,Feng:1999fw,Feng:1999zv}. In our brane
setup obtained from T-duality, however, the NS5-branes are smeared
while in the brane box models all the fivebranes are localized.
Nevertheless this does not affect the main picture or the fact
that we are dealing with a (3+1 dimensional) conformal field
theory.

The brane setup of the brane box model is as follows
\cite{Hanany-Uranga,Hanany:1997tb}:

$\bullet$ $N$ D5-branes along 012345.

$\bullet$ $p$ NS5-branes along 012346, and uniformly distributed on
the $x^5$ direction.

$\bullet$ $q$ NS5-branes along 012357, and uniformly distributed
along the $x^4$ direction.
\\
The $x^4$ and $x^5$ directions are periodically identified with
radii $R_4$ and $R_5$. The 45 plane is then like a two dimensional
lattice with $p \times q$ sites.\footnote{As mentioned earlier in
section \ref{the-lattice}, this torus may be viewed as a
latticized fuzzy torus, and since $p=q$ the fuzziness is
$\Theta=1/p$.} The size of the unit cell on the lattice is
$\frac{R_4R_5}{pq}$.

The low energy effective theory living on the above intersecting
brane setup is a 3+1 dimensional $U(N)^{pq}$ gauge theory with $A, B$ and $C$
bi-fundamentals corresponding to open strings stretched between
segments of the D5-branes, i.e. they are the links on the lattice,
exactly as explained earlier. The intersecting brane system is 1/8
BPS and preserves 4 supercharges; therefore, the low energy
effective theory is an $\cN=1$ SYM theory. The rotation in the
$89$ plane then corresponds to the $U(1)$ $R$-symmetry of the
theory. The couplings of all the $pq$ $U(N)$ factors are equal, as
we have chosen a uniform distribution of NS5-branes. This coupling
is equal to the coupling of the
$5+1$ dimensional $U(N)$ SYM on the $N$ D5-branes divided by the
volume of each cell: $g^2_{5+1\ YM}\times
\frac{pq}{R_4R_5}=g^2_{3+1\ YM}\ pq$; $g^2_{3+1\ YM}=g_s$
\cite{Hanany-Uranga}. This is another way of stating
eq.\eqref{N=1-coupling}. We see that the explicit dependence on
the size of the torus, or the lattice spacing, drops out; this is
a sign of the conformal symmetry at the level of the $3+1$
dimensional theory.  As a result, the lattice spacing remains an
arbitrary parameter. The superpotential of this model, which has a
natural appearance on the $2d$ oriented triangle lattice ({\it
cf.} \eqref{superpotential}) may also be read from the brane setup
\cite{Hanany:1997tb}.

The brane box model provides a more geometric view of the $2+1$
dimensional (conformal) lattice we have developed in previous
sections. Moreover, it makes closer contact with string theory.
In this setup one may interpret the $2+1$ latticized string theory
as a discretized version of a specific sector of the $\cN=(1,1)$
little string theory living on the type IIB fivebranes. From this
viewpoint, the time direction on the lattice theory is identified
with the time direction along the branes. The brane box model also
suggests that if we include the site variables, i.e. the
vector multiplets of the $3+1$ dimensional $\cN=1$ theory, we should obtain a
six dimensional string theory with two directions on a fuzzy two-torus.

\section{Higgsed Phase}
\label{Higgs}

So far we have studied the $\cN=1$ gauge theory at its conformal
fixed point (line). This corresponds to the case where the
bi-fundamentals have zero VEV. It is possible to give a non-zero
VEV to $A$'s, $B$'s and $C$'s in a such a way that we preserve the
supersymmetry and hence the $U(1)$ R-symmetry. This would,
however,  break the conformal symmetry. {}From the gravity
viewpoint discussed in the previous section this corresponds to
taking the stack of  $N$ D3-branes away from the orbifold fixed
point. From the two dimensional lattice point of view, however,
this corresponds to introducing a specific length scale on the two
dimensional lattice. In other words, the spacing of the lattice
now depends on the VEV's.

In this section we consider such Higgsing and study the special
case that all of the $3pq$ bi-fundamentals acquire  non-zero, but
equal VEV's.{} First  we check that this leads to a $2+1$ dimensional
effective $U(N)$ lattice gauge theory, now with fixed lattice
spacing and next discuss its connection to deconstruction
\cite{Arkani-Hamed:2001ie}.

\subsection{Lattice Gauge Theory Picture}
\label{LGT}

What we have is a three-dimensional Euclidean lattice gauge
theory in a temporal gauge where the timelike links have had their
field values set to zero. This is a discretized Hamiltonian
lattice gauge theory. The lattice action of this system is just
the plaquette action we have already presented. The operators we
have been discussing are literally Wilson loops in this picture.
A three dimensional lattice gauge theory has two fixed
points: the IR fixed point a  UV fixed point. The UV fixed point
is a trivial fixed point about which the theory is free. The IR
fixed point is a non-trivial one at which the theory flows to
an interacting three dimensional conformal field theory. In the IR our
lattice gauge theory obviously would look like a continuum theory
as we cannot probe the discreteness of the lattice. In
fact, one should be more careful, because we are working on a
torus. The above then becomes exact only for large $p,q$.

Our viewpoint then provides a relation between
the transfer matrix already presented and a three-dimensional
Euclidean lattice gauge theory.  We take the view that the expansion
on the lattice in terms of plaquettes
can be reinterpreted as the strong coupling expansion for a lattice
gauge theory, in effect defining its Hamiltonian. The lattice gauge theory
is formulated in the language of Hamiltonian lattice gauge theory,
relying on a continuum Euclidean time direction, with the
spacial directions latticized\footnote{This approach explicitly breaks
symmetries relating
space and time but make the spectrum of the theory more transparent.
The Hamiltonian formulation of lattice gauge theory can be derived
from the more common formulation in which time is Euclideanized and discrete
by introducing a different lattice spacing (and coupling) for the time direction and studying the limit as this spacing is taken to vanish, adjusting the couplings appropriately.}
Consider this formulation in temporal gauge, where all vector fields in the time
direction are set to zero, $A_0=0$. Then the Hamiltonian depends only
on space-like components of the vector fields, together with the
conjugate momenta. Our interaction
plaquettes then provide a strong coupling expansion
of such a Hamiltonian, thus providing us with an explicit
construction of the lattice gauge theory. \footnote{For some other relevant constructions see
\cite{Mithat}.}

\subsection{Relation to Deconstruction}
\label{deconstruction}

The four-dimensional $\mathcal{N}=1$ superconformal field theory
can be Higgsed down to the diagonal subgroup of $\supq$ by giving
vacuum expectation value to the link variables, with all link
variables in a particular direction taking on the same VEV
\cite{Arkani-Hamed:2001ie}.\footnote{Similar ideas have appeared much earlier in \cite{Halpern}.} This leads to a picture of
deconstructed extra dimensions: at intermediate energies, the
dynamics of this theory becomes that of a non-chiral
six-dimensional theory with $\mathcal{N}=(1,1)$ supersymmetry,
where the lattice directions of the Moose, which has been
toroidally identified, is a discretization of two space-like
directions, compactified on a torus. There are two energy scales
which determine the range within which this higher-dimensional
dynamics emerges, an inverse effective lattice spacing $a^{-1}$,
and the inverse size of the compact directions
$R^{-1}$.\footnote{For simplicity we consider the case where the
VEV's are chosen such that the lattice spacing and radii in the
two compact latticized direction are equal. The more general case
can also be considered, but the essential physics is the same. The
complex structure modulus and radii of the torus are determined by
the specific VEV's assigned to the three directions $A, B, C$, as
well as the four-dimensional gauge coupling.} The inverse lattice
spacing is equal to VEV of the link variables times the
four-dimensional gauge coupling of the individual $U(N)$ theories
(all taken to be the same throughout the paper). From the lattice
picture of the Moose, it is clear that $R \approx p a$, with $p$
the number of gauge theory nodes in one Moose direction. The range
of energies where this picture is a good description of the
physics is intermediate between the inverse lattice spacing on the
high side, and the inverse of the compactification radii on the
low side. In this intermediate regime, there are massive
excitations which are interpreted as Kaluza-Klein modes of the
compactified six-dimensional theory. At energies below the inverse
radii, the KK tower is no longer excited, and we recover a
four-dimensional effective description. The ultraviolet behaviour
of the six-dimensional theory is regulated by that fact that at
energies above $a^{-1}$, the physics reverts back to that of the
original four-dimensional conformal theory. Notice also that the
six-dimensional theory is not conformally invariant, since the
gauge coupling outside four-dimensional is dimensionful, being set
by the scale of the VEV's.

In this scenario, the lattice plaquettes we found in the previous
section arise from the discretization of terms in the
six-dimensional field theory potential.

Finally, a scaling limit can be taken, with the lattice spacing
appropriately scaled to zero and the four-dimensional coupling
taken to infinity, which yields an interacting continuum
six-dimensional theory describing $(1,1)$ little string theory
\cite{Arkani-Hamed:2001ie}.

Our discussion in this paper was mainly related to the
superconformal point in the moduli space of the $\mathcal{N}=1$
theory, where all VEV's vanish, and so is not directly related to
deconstruction. However, there is a relation to six-dimensional
theories, and the intersecting type IIB fivebranes which we
touched on in section \ref{branebox}.

\section{Summary and Discussion}
\label{summary}

In this paper we have considered the $\cN=1$ superconformal
$U(N)^{pq}$ gauge theory with $3pq$ bi-fundamental chiral
multiplets. We have explored the fact that all the information
about this theory, namely its superpotential and its gauge
invariant \opt s, can be summarized on a $2d$ oriented triangle
lattice. In this lattice picture bi-fundamentals appear as the
link variables and vector gauge fields as site variables. Although
we mainly focused on the bosonic part of the chiral multiplets, as
we briefly mentioned this lattice can be thought of as a
``super-lattice'' where links represent the full chiral
multiplet and not just the scalar field.

We focused on the computation of the dilatation \opt\ of this
theory and gave an explicit representation of the dilatation \opt\
at one-loop planar level on the lattice. Using this information we
can then compute the one-loop anomalous dimension of any (gauge
invariant) \opt\ of the theory. However, for simplicity we
focused on the \opt s constructed only from the bi-fundamentals. The
gauge invariant \opt s in this sector are the oriented closed
loops on the $2d$ lattice. We have shown that the dilatation \opt\
acts like a Laplacian (plus some ``contact terms'') on the $2d$
lattice. As such, the gauge invariant \opt s may be thought of as
states in the configuration space of the $2+1$ dimensional
oriented closed string theory, with a latticized target space.
This $2d$ target space, for finite $p, q$, is a $2d$ fuzzy torus
with $pq$ points on it and with noncommutativity parameter
$\Theta=\frac{{\rm gcd(p,q)}}{pq}$.

In another interpretation, the dilatation \opt\ which is the
Hamiltonian of the gauge theory on $R\times S^3$, can also be
taken as the transfer matrix for a $2d$ statistical mechanical
system on the $2d$ lattice. Recall that the $\cN=1$ theory we have
considered arises from the $\cN=4$ SYM via orbifolding and the
fact that the latter is related to a one dimensional spin chain
system, which is an integrable model. As we argued, the $2d$
statistical mechanical
system, corresponding to the one loop planar dilatation
\opt\ restricted to the (anti-)holomorphic sector of the $\cN=1$
gauge theory \opt s, is also integrable; it is the $SU(3)$
anti-ferromagnetic spin chain. One may then try to define the
orbifolding on the statistical mechanical model directly without
invoking
the gauge theory in such a way that integrability is preserved.
As we have seen explicitly, this orbifolding can relate a higher
dimensional statistical mechanical system to a lower dimensional one, which
is generically more tractable. Crystalizing and elaborating on
this idea is of course of great interest.

We have shown that only the F-terms contribute to the anomalous
dimensions of the \opt s in the (anti-)holomorphic sector and the
F-terms are the contributions of the superpotential to the action.
On the other hand the integrable $SU(3)$ spin chain structure is
inferred from the specific form of the action. Therefore, the
integrability should be closely related to the form of the
superpotential. Are there new non-renormalization theorems
resulting from  the integrability of the lattice theory? It is
desirable to make this connection clearer and formulate possible
new non-renormalization theorems.

We focused on the one loop planar dilatation \opt\ restricted
to the holomorphic \opt s, corresponding to a $SU(3)$ spin chain
with nearest neighbor interactions. As the holomorphic sector
closes onto itself, including the higher loop effects, we expect
to still find an $SU(3)$ spin chain but now with longer range
interactions. As in the $SO(6)$ spin chain of the $\cN=4$ theory
one would still expect that the integrability carries over (to all
loop order). Providing arguments for this expectation is an
interesting line for future work.

We also discussed the theory away from the conformal fixed line,
in the Higgsed phase. We argued that in a specific Higgsing,
where all the VEV's of bi-fundamentals are taken equal, the
dilatation \opt\ on the space of all gauge invariant \opt s made out of
bi-fundamentals is indeed equivalent to a 2+1 dimensional $U(N)$
lattice gauge theory. It is interesting to check if the
integrability extends beyond the conformal fixed line and to the
Higgsed phase; if this is true (there have already been conjectures
in this direction \cite{Wang:2003cu})  one can then directly argue
for the integrability of the $3d$ lattice gauge theory.

As we briefly discussed, when including the gauge fields as well as
the bi-fundamentals, the effective  theory becomes a
six-dimensional string theory, a little string theory on a
six-dimensional space in which two directions lie on a fuzzy torus.
The integrability in the holomorphic sector then implies
integrability of a sector of this little string theory. An
interesting open question is whether the integrability arguments
in the holomorphic sector can be extended to the full little
string theory.

In our example where we started with $\cN=4, D=4$ SYM we can at
most have two dimensional quivers which preserve conformal
symmetry. For $\cN=8, D=3$ superconformal theories, however, we
have the possibility of three dimensional quivers, leading to cubic
lattices. In this way one may be able to do $3+1$ dimensional
(lattice) gauge theory analysis via the $D=3$ SCFT. This is an
interesting direction for further studies. On a $3d$ lattice
the configurations can be labeled by all different (orientable)
closed $2d$ surfaces. In this viewpoint one would expect that a
$3d$ lattice theory leads to a membrane theory with a $3+1$
dimensional target space.

Here we mainly focused on the ``holomorphic'' \opt s of the
$\cN=1,\ D=4$ gauge theory. For this sector, up to some
subtleties, one may use a representation on the dual hexagonal
lattice. The orientation on the triangles of the original lattice
then translates into a positive or negative charge assignment to
the sites on the hexagonal lattice. Using the hexagonal dual
lattice and this charge assignment one can then find a one-to-one
relation between the holomorphic closed loop on the triangle
lattice and an ``Ising type'' configurations on the hexagonal
lattice. One can consider closed loops on the triangle lattice in
which a specific link is repeated several times. For these cases
one should extend the above charge assignment from just $\pm 1$ to
$\pm 1$, $\pm 2$, $\pm 3$, $\cdots$. (Note that this charge
assignment is different from the $R$-charge of the \opt .)
Therefore, at least at the level of the configuration space the
holomorphic sector of the gauge theory is mapped onto a Potts
model, rather than Ising. It is then straightforward to re-write
the action of the dilatation \opt\ on this hexagonal lattice,
which appears to be a nearest neighbor Hamiltonian corresponding
to a $Q$-state Potts model (with large $Q$). Here we did not
address the theory from the hexagonal dual lattice viewpoint.
There is an obvious interest in the further exploration of this
viewpoint. For example whether one can use the integrability
structures more apparent in the gauge theory language to study the
integrability and phase structure of $3d$ Ising/Potts model
(recall also the previous paragraph.) There have also been papers discussing the relation been the brane box models, dimer models and Ising on the hexagonal lattice \cite{dimer}. It is interesting to study the implication of the integrability we discussed here to these cases.

\begin{figure}[ht]
\centering
\epsfig{figure=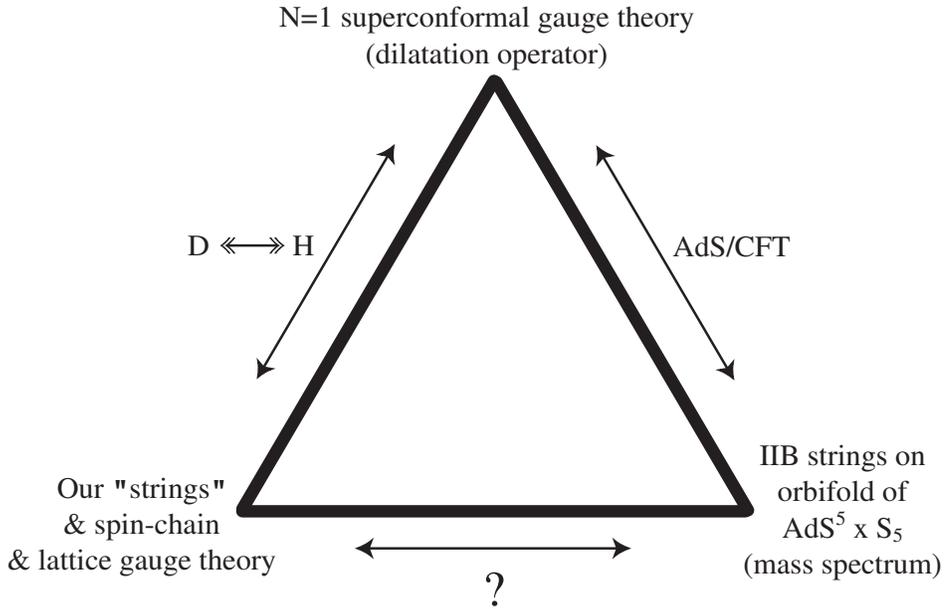, width=372pt, height=234pt}
\begin{center}
\caption{The dualities.}
\label{final}
\end{center}
\end{figure}

Our results are captured succinctly as legs of a triangle
(see figure \ref{final}); the
AdS/CFT duality relating the $\mathcal{N}=1$ gauge theory and orbifolds
of string theory provides one leg; the main focus of this paper has been
on a leg relating the gauge theory to a new 2+1-dimensional ``string
theory''; this is predicated on the identification of the gauge theory's
dilatation operator as the Hamiltonian of the string theory.
The third leg of the triangle, relating the ten-dimensional strings
on the orbifold space and the 2+1-dimensional theory described in this
paper is open to future exploration.

\section*{Acknowledgments}

It is a pleasure to thank Scott Thomas for his collaboration in
some stages of this work. We would like to thank John McGreevy,
Jan Plefka and Steve Shenker for useful discussions. The work of
D.S. was supported by the US National Science Foundation under
grant PHY02-44728. The work of M. M. Sh-J. was supported in part
by NSF grant PHY-9870115 and in part by funds from the Stanford
Institute for Theoretical Physics.

\appendix

\section{Conventions}
\label{appendix-A}

We present in this appendix our conventions and notation.
$SU(N)$ traces satisfy
\be \label{traces}
  tr(T^a T^b) = C(r) \delta^{ab}
\ee%
 with $C(r)$ the Dynkin index of the representation for the
fields, equal to $N$ for the adjoint representation. Writing the
adjoint fields of $U(N)$ as products of fundamental and
anti-fundamentals allows us to write free propagators with the
index structure (dropping the obvious space-time dependance): %
\be
  \Big< \phi^i_j \phi^k_l \Big>_0 \propto
  \delta^i_l \delta^k_j
\ee%
for $U(N)$. For $SU(N)$ there is an extra term arising from the
fact that all the generators are traceless. Subtracting the extra
trace part gives \be
  \Big< \phi^i_j \phi^k_l \Big>_0 \propto
  \left( \delta^i_l \delta^k_j - \frac{1}{N} \delta^i_j \delta^k_l \right)
\ee
The extra term is subleading in the large $N$ limit, but in any case, because of the commutators in
\eqref{dilatation-real} we are free to use the $U(N)$ rules.

We also have the useful identity
\be
  tr \left( T^a [ T^b , T^c ] \right) = i C(r) f^{abc}
\ee
where $f^{abc}$ are the structure constants. Note also that
\be
  \epsilon_{ijk} \epsilon_{ilm} = \delta^{jl} \delta^{km} - \delta^{jm} \delta^{kl}
\ee
summed over $i=1,2,3$.

Following these conventions, the $\mathcal{N}=4$ Lagrangian can be written in $\mathcal{N}=1$ language as follows:
\be \label{N4-action}
\begin{aligned}
  S \: = \:
  \frac{1}{C(r)} \: tr \: & \Bigg[
  \int d^4 \theta \: \:
  e^{-g_{YM} V} \: \bar{\Phi}_i \: e^{g_{YM} V} \: \Phi_i \\
  &+ \:
  \frac{1}{16 g_{YM}^2} \left(
  \int d^2 \theta \: \: W^\alpha W_\alpha \: + \:
  \int d^2 \bar{\theta} \: \: \bar{W}_{\dot{\alpha}} \bar{W}^{\dot{\alpha}}
  \right) \\
  &+ \: i \: g_{YM} \frac{\sqrt{2}}{3 !} \epsilon_{ijk}
  \left(
  \int d^2 \theta \: \: \Phi_i [ \Phi_j , \Phi_ k ] \: + \:
  \int d^2 \bar{\theta} \: \: \bar{\Phi}_i [ \bar{\Phi}_j , \bar{\Phi}_k ]
  \right)
  \Bigg]
\end{aligned}
\ee The $\Phi_i$ are three chiral superfields and $V$ is a vector
superfield, all transforming in the adjoint representation of
$U(Npq)$.\footnote{We follow the conventions of Wess and Bagger
\cite{Wess:1992cp}.}. The correct dependence on the coupling for
the gauge kinetic terms is established after taking $V \rightarrow
2 g_{YM} V$ in the second term.

When writing the $\mathcal{N}=4$ action in $\mathcal{N}=1$ language,
the F-terms and D-terms are
\begin{equation}
  \mathcal{L}_D \: = \: \frac{-1}{2 g_{YM}^2} \Tr \left(
  [ A , \bar{A} ] \: + \: [ B , \bar{B} ] \: + \: [ C , \bar{C} ]
  \right)^2
\end{equation}
with bars denoting conjugation, and
\begin{equation}
  \mathcal{L}_F \: = \: \frac{1}{g_{YM}^2} \Tr \left(
  | [ A , B ] |^2 \: + \: | [ A , C ] |^2 \: + \: \| [ B , C ] |^2
  \right)
\end{equation}
The trace here is taken over $p \times q \times N$ by $p \times q \times N$ matrices.
The F-terms arise from a superpotential of the form $Tr(ABC)$.
As in the parent $\mathcal{N}=4$ theory,
the gauge field loops, scalar self energies and D-term contributions cancel against each other. This is
responsible for making the straight line operators BPS, as discussed in section \ref{transfer-matrix}.

The orbifolding in the gauge theory acts as a projection on these matrices.
The square of the commutator terms in the F-terms appear as
\begin{equation}
  | [ A , B ] |^2 \: = \: [ \bar{B} \: , \: \bar{A} ] [ A \: , \: B ]
\end{equation}

A useful shorthand notation for capturing field contractions is to introduce
variations with respect to the fields as follows
\begin{equation}
  \delta_m = \frac{\delta}{\delta \phi_m}
\end{equation}
The six real scalar fields $\phi_m$ are matrix valued $\phi_m =
\phi_m^a T^a$, with $T^a$ being the generators of $U(Npq)$. This
definition is motivated by analogy to Wick contractions in the
form of free field propagators.

We find it convenient to work in terms of the complex scalars appearing in the three chiral superfields, letting
\begin{equation}
  \Phi_I = \frac{1}{\sqrt{2}} \left( \phi_{2I-1} + i \phi_{2I} \right) \\
\end{equation}
with $I=1,2,3$ and $\{\Phi_I\}=A,B,C$.
Then the variations with respect to these complex fields are defined as
\begin{equation}
  \Delta_I =
    \frac{1}{\sqrt{2}} \left( \frac{\delta}{\delta \phi_{2I-1}}
    - i \frac{\delta}{\delta \phi_{2I}} \right)
\end{equation}
and the derivatives with respect to the conjugate fields
are
\begin{equation}
  \bar{\Delta}_I =
    \frac{1}{\sqrt{2}} \left( \frac{\delta}{\delta \phi_{2I-1}}
    + i \frac{\delta}{\delta \phi_{2I}} \right)
\end{equation}
Since the Wick contractions of the scalars in non-zero only when
we contract a field and its conjugate, we require that the
variation $\Delta_I$ transform in the same way as $\bar{\Phi}_I$,
i.e.\! in the conjugate representation, and therefore define
$\Delta_I = \Delta_I^a \bar{T}^a$. Note that (as explained in
section \ref{gauge-theory-orbifold}) the generators $T^a$ on the
covering space are not hermitian. Note also that derivatives act
as annihilation operators, and the fields and derivatives satisfy
a creation-annihilation algebra. The example in Appendix B
clarifies these points.

\section{Orbifolding Example}
\label{appendix-B}

We give an explicit example of the orbifolding in the gauge
theory, where we take $p=q=3$, for which there are $9$ lattice
sites. The generators of the symmetry for the theory living on the
covering space is the direct product of the generators of the
original $U(9N)$ symmetry with the generators of $\mathbb{Z}_3
\times \mathbb{Z}_3$ in the regular representation, leading to the
generators\footnote{$T^\alpha_{i,j}$, etc. are the generators of
$U(N)$. They are hermitian, so $\bar{T}^\alpha_{i,j} \equiv
(T^\alpha)^\dagger_{i,j} = T^\alpha_{i,j}$.}
\begin{equation} \label{TA33}
T^{A \alpha} \: = \:
  \left(
  \begin{matrix}
    0 \: \: & \: \: T^{\alpha}_{1,1} \: \: & \: \: 0 \: \: & \: \: 0 \: \: & \: \: 0 \: \: & \: \: 0
    \: \: & \: \: 0 \: \: & \: \: 0 \: \: & \: \: 0 \\
    0 \: \: & \: \: 0 \: \: & \: \: T^{\alpha}_{1,2} \: \: & \: \: 0 \: \: & \: \: 0 \: \: & \: \: 0
    \: \: & \: \: 0 \: \: & \: \: 0 \: \: & \: \: 0 \\
    T^{\alpha}_{1,3} \: \: & \: \: 0 \: \: & \: \: 0 \: \: & \: \: 0 \: \: & \: \: 0 \: \: & \: \: 0
    \: \: & \: \: 0 \: \: & \: \: 0 \: \: & \: \: 0 \\
    0 \: \: & \: \: 0 \: \: & \: \: 0 \: \: & \: \: 0 \: \: & \: \: T^{\alpha}_{2,1} \: \: & \: \: 0
    \: \: & \: \: 0 \: \: & \: \: 0 \: \: & \: \: 0 \\
    0 \: \: & \: \: 0 \: \: & \: \: 0 \: \: & \: \: 0 \: \: & \: \: 0 \: \: & \: \: T^{\alpha}_{2,2}
    \: \: & \: \: 0 \: \: & \: \: 0 \: \: & \: \: 0 \\
    0 \: \: & \: \: 0 \: \: & \: \: 0 \: \: & \: \: T^{\alpha}_{2,3} \: \: & \: \: 0 \: \: & \: \: 0
    \: \: & \: \: 0 \: \: & \: \: 0 \: \: & \: \: 0 \\
    0 \: \: & \: \: 0 \: \: & \: \: 0 \: \: & \: \: 0 \: \: & \: \: 0 \: \: & \: \: 0
    \: \: & \: \: 0 \: \: & \: \: T^{\alpha}_{3,1} \: \: & \: \: 0 \\
    0 \: \: & \: \: 0 \: \: & \: \: 0 \: \: & \: \: 0 \: \: & \: \: 0 \: \: & \: \: 0
    \: \: & \: \: 0 \: \: & \: \: 0 \: \: & \: \: T^{\alpha}_{3,2} \\
    0 \: \: & \: \: 0 \: \: & \: \: 0 \: \: & \: \: 0 \: \: & \: \: 0 \: \: & \: \: 0
    \: \: & \: \: T^{\alpha}_{3,3} \: \: & \: \: 0 \: \: & \: \: 0
  \end{matrix}
  \right)
\end{equation}
\begin{equation} \label{TB33}
T^{B \alpha} \: = \:
  \left(
  \begin{matrix}
    0 \: \: & \: \: 0 \: \: & \: \: 0 \: \: & \: \: 0 \: \: & \: \: 0 \: \: & \: \: 0
    \: \: & \: \: 0 \: \: & \: \: 0 \: \: & \: \: T^{\alpha}_{1,1} \\
    0 \: \: & \: \: 0 \: \: & \: \: 0 \: \: & \: \: 0 \: \: & \: \: 0 \: \: & \: \: 0
    \: \: & \: \: T^{\alpha}_{1,2} \: \: & \: \: 0 \: \: & \: \: 0 \\
    0 \: \: & \: \: 0 \: \: & \: \: 0 \: \: & \: \: 0 \: \: & \: \: 0 \: \: & \: \: 0
    \: \: & \: \: 0 \: \: & \: \: T^{\alpha}_{1,3} \: \: & \: \: 0 \\
    0 \: \: & \: \: 0 \: \: & \: \: T^{\alpha}_{2,1} \: \: & \: \: 0 \: \: & \: \: 0 \: \: & \: \: 0
    \: \: & \: \: 0 \: \: & \: \: 0 \: \: & \: \: 0 \\
    T^{\alpha}_{2,2} \: \: & \: \: 0 \: \: & \: \: 0 \: \: & \: \: 0 \: \: & \: \: 0 \: \: & \: \: 0
    \: \: & \: \: 0 \: \: & \: \: 0 \: \: & \: \: 0 \\
    0 \: \: & \: \: T^{\alpha}_{2,3} \: \: & \: \: 0 \: \: & \: \: 0 \: \: & \: \: 0 \: \: & \: \: 0
    \: \: & \: \: 0 \: \: & \: \: 0 \: \: & \: \: 0 \\
    0 \: \: & \: \: 0 \: \: & \: \: 0 \: \: & \: \: 0 \: \: & \: \: 0 \: \: & \: \: T^{\alpha}_{3,1}
    \: \: & \: \: 0 \: \: & \: \: 0 \: \: & \: \: 0 \\
    0 \: \: & \: \: 0 \: \: & \: \: 0 \: \: & \: \: T^{\alpha}_{3,2} \: \: & \: \: 0 \: \: & \: \: 0
    \: \: & \: \: 0 \: \: & \: \: 0 \: \: & \: \: 0 \\
    0 \: \: & \: \: 0 \: \: & \: \: 0 \: \: & \: \: 0 \: \: & \: \: T^{\alpha}_{3,3} \: \: & \: \: 0
    \: \: & \: \: 0 \: \: & \: \: 0 \: \: & \: \: 0
  \end{matrix}
  \right)
\end{equation}
\begin{equation} \label{TC33}
T^{C \alpha} \: = \:
  \left(
  \begin{matrix}
    0 \: \: & \: \: 0 \: \: & \: \: 0 \: \: & \: \: T^{\alpha}_{1,1} \: \: & \: \: 0 \: \: & \: \: 0
    \: \: & \: \: 0 \: \: & \: \: 0 \: \: & \: \: 0 \\
    0 \: \: & \: \: 0 \: \: & \: \: 0 \: \: & \: \: 0 \: \: & \: \: T^{\alpha}_{1,2} \: \: & \: \: 0
    \: \: & \: \: 0 \: \: & \: \: 0 \: \: & \: \: 0 \\
    0 \: \: & \: \: 0 \: \: & \: \: 0 \: \: & \: \: 0 \: \: & \: \: 0 \: \: & \: \: T^{\alpha}_{1,3}
    \: \: & \: \: 0 \: \: & \: \: 0 \: \: & \: \: 0 \\
    0 \: \: & \: \: 0 \: \: & \: \: 0 \: \: & \: \: 0 \: \: & \: \: 0 \: \: & \: \: 0
    \: \: & \: \: T^{\alpha}_{2,1} \: \: & \: \: 0 \: \: & \: \: 0 \\
    0 \: \: & \: \: 0 \: \: & \: \: 0 \: \: & \: \: 0 \: \: & \: \: 0 \: \: & \: \: 0
    \: \: & \: \: 0 \: \: & \: \: T^{\alpha}_{2,2} \: \: & \: \: 0 \\
    0 \: \: & \: \: 0 \: \: & \: \: 0 \: \: & \: \: 0 \: \: & \: \: 0 \: \: & \: \: 0
    \: \: & \: \: 0 \: \: & \: \: 0 \: \: & \: \: T^{\alpha}_{2,3} \\
    T^{\alpha}_{3,1} \: \: & \: \: 0 \: \: & \: \: 0 \: \: & \: \: 0 \: \: & \: \: 0 \: \: & \: \: 0
    \: \: & \: \: 0 \: \: & \: \: 0 \: \: & \: \: 0 \\
    0 \: \: & \: \: T^{\alpha}_{3,2} \: \: & \: \: 0 \: \: & \: \: 0 \: \: & \: \: 0 \: \: & \: \: 0
    \: \: & \: \: 0 \: \: & \: \: 0 \: \: & \: \: 0 \\
    0 \: \: & \: \: 0 \: \: & \: \: T^{\alpha}_{3,3} \: \: & \: \: 0 \: \: & \: \: 0 \: \: & \: \: 0
    \: \: & \: \: 0 \: \: & \: \: 0 \: \: & \: \: 0
  \end{matrix}
  \right)
\end{equation}

The conjugate representation is generated by
\begin{equation} \label{TAbar33}
\bar{T}^{A \alpha} \: = \:
  \left(
  \begin{matrix}
    0 \: \: & \: \: 0 \: \: & \: \: \bar{T}^{\alpha}_{1,3} \: \: & \: \: 0 \: \: & \: \: 0 \: \: & \: \: 0
    \: \: & \: \: 0 \: \: & \: \: 0 \: \: & \: \: 0 \\
    \bar{T}^{\alpha}_{1,1} \: \: & \: \: 0 \: \: & \: \: 0 \: \: & \: \: 0 \: \: & \: \: 0 \: \: & \: \: 0
    \: \: & \: \: 0 \: \: & \: \: 0 \: \: & \: \: 0 \\
    0 \: \: & \: \: \bar{T}^{\alpha}_{1,2} \: \: & \: \: 0 \: \: & \: \: 0 \: \: & \: \: 0 \: \: & \: \: 0
    \: \: & \: \: 0 \: \: & \: \: 0 \: \: & \: \: 0 \\
    0 \: \: & \: \: 0 \: \: & \: \: 0 \: \: & \: \: 0 \: \: & \: \: 0 \: \: & \: \: \bar{T}^{\alpha}_{2,3}
    \: \: & \: \: 0 \: \: & \: \: 0 \: \: & \: \: 0 \\
    0 \: \: & \: \: 0 \: \: & \: \: 0 \: \: & \: \: \bar{T}^{\alpha}_{2,1} \: \: & \: \: 0 \: \: & \: \: 0
    \: \: & \: \: 0 \: \: & \: \: 0 \: \: & \: \: 0 \\
    0 \: \: & \: \: 0 \: \: & \: \: 0 \: \: & \: \: 0 \: \: & \: \: \bar{T}^{\alpha}_{2,2} \: \: & \: \: 0
    \: \: & \: \: 0 \: \: & \: \: 0 \: \: & \: \: 0 \\
    0 \: \: & \: \: 0 \: \: & \: \: 0 \: \: & \: \: 0 \: \: & \: \: 0 \: \: & \: \: 0
    \: \: & \: \: 0 \: \: & \: \: 0 \: \: & \: \: \bar{T}^{\alpha}_{3,3} \\
    0 \: \: & \: \: 0 \: \: & \: \: 0 \: \: & \: \: 0 \: \: & \: \: 0 \: \: & \: \: 0
    \: \: & \: \: \bar{T}^{\alpha}_{3,1} \: \: & \: \: 0 \: \: & \: \: 0 \\
    0 \: \: & \: \: 0 \: \: & \: \: 0 \: \: & \: \: 0 \: \: & \: \: 0 \: \: & \: \: 0
    \: \: & \: \: 0 \: \: & \: \: \bar{T}^{\alpha}_{3,2} \: \: & \: \: 0
  \end{matrix}
  \right)
\end{equation}
\begin{equation} \label{TBbar33}
\bar{T}^{B \alpha} \: = \:
  \left(
  \begin{matrix}
    0 \: \: & \: \: 0 \: \: & \: \: 0 \: \: & \: \: 0 \: \: & \: \: \bar{T}^{\alpha}_{2,2} \: \: & \: \: 0
    \: \: & \: \: 0 \: \: & \: \: 0 \: \: & \: \: 0 \\
    0 \: \: & \: \: 0 \: \: & \: \: 0 \: \: & \: \: 0 \: \: & \: \: 0 \: \: & \: \: \bar{T}^{\alpha}_{2,3}
    \: \: & \: \: 0 \: \: & \: \: 0 \: \: & \: \: 0 \\
    0 \: \: & \: \: 0 \: \: & \: \: 0 \: \: & \: \: \bar{T}^{\alpha}_{2,1} \: \: & \: \: 0 \: \: & \: \: 0
    \: \: & \: \: 0 \: \: & \: \: 0 \: \: & \: \: 0 \\
    0 \: \: & \: \: 0 \: \: & \: \: 0 \: \: & \: \: 0 \: \: & \: \: 0 \: \: & \: \: 0
    \: \: & \: \: 0 \: \: & \: \: \bar{T}^{\alpha}_{3,2} \: \: & \: \: 0 \\
    0 \: \: & \: \: 0 \: \: & \: \: 0 \: \: & \: \: 0 \: \: & \: \: 0 \: \: & \: \: 0
    \: \: & \: \: 0 \: \: & \: \: 0 \: \: & \: \: \bar{T}^{\alpha}_{3,3} \\
    0 \: \: & \: \: 0 \: \: & \: \: 0 \: \: & \: \: 0 \: \: & \: \: 0 \: \: & \: \: 0
    \: \: & \: \: \bar{T}^{\alpha}_{3,1} \: \: & \: \: 0 \: \: & \: \: 0 \\
    0 \: \: & \: \: \bar{T}^{\alpha}_{1,2} \: \: & \: \: 0 \: \: & \: \: 0 \: \: & \: \: 0 \: \: & \: \: 0
    \: \: & \: \: 0 \: \: & \: \: 0 \: \: & \: \: 0 \\
    0 \: \: & \: \: 0 \: \: & \: \: \bar{T}^{\alpha}_{1,3} \: \: & \: \: 0 \: \: & \: \: 0 \: \: & \: \: 0
    \: \: & \: \: 0 \: \: & \: \: 0 \: \: & \: \: 0 \\
    \bar{T}^{\alpha}_{1,1} \: \: & \: \: 0 \: \: & \: \: 0 \: \: & \: \: 0 \: \: & \: \: 0 \: \: & \: \: 0
    \: \: & \: \: 0 \: \: & \: \: 0 \: \: & \: \: 0
  \end{matrix}
  \right)
\end{equation}
\begin{equation} \label{TCbar33}
\bar{T}^{C \alpha} \: = \:
  \left(
  \begin{matrix}
    0 \: \: & \: \: 0 \: \: & \: \: 0 \: \: & \: \: 0 \: \: & \: \: 0 \: \: & \: \: 0
    \: \: & \: \: \bar{T}^{\alpha}_{3,1} \: \: & \: \: 0 \: \: & \: \: 0 \\
    0 \: \: & \: \: 0 \: \: & \: \: 0 \: \: & \: \: 0 \: \: & \: \: 0 \: \: & \: \: 0
    \: \: & \: \: 0 \: \: & \: \: \bar{T}^{\alpha}_{3,2} \: \: & \: \: 0 \\
    0 \: \: & \: \: 0 \: \: & \: \: 0 \: \: & \: \: 0 \: \: & \: \: 0 \: \: & \: \: 0
    \: \: & \: \: 0 \: \: & \: \: 0 \: \: & \: \: \bar{T}^{\alpha}_{3,3} \\
    \bar{T}^{\alpha}_{1,1} \: \: & \: \: 0 \: \: & \: \: 0 \: \: & \: \: 0 \: \: & \: \: 0 \: \: & \: \: 0
    \: \: & \: \: 0 \: \: & \: \: 0 \: \: & \: \: 0 \\
    0 \: \: & \: \: \bar{T}^{\alpha}_{1,2} \: \: & \: \: 0 \: \: & \: \: 0 \: \: & \: \: 0 \: \: & \: \: 0
    \: \: & \: \: 0 \: \: & \: \: 0 \: \: & \: \: 0 \\
    0 \: \: & \: \: 0 \: \: & \: \: \bar{T}^{\alpha}_{1,3} \: \: & \: \: 0 \: \: & \: \: 0 \: \: & \: \: 0
    \: \: & \: \: 0 \: \: & \: \: 0 \: \: & \: \: 0 \\
    0 \: \: & \: \: 0 \: \: & \: \: 0 \: \: & \: \: \bar{T}^{\alpha}_{2,1} \: \: & \: \: 0 \: \: & \: \: 0
    \: \: & \: \: 0 \: \: & \: \: 0 \: \: & \: \: 0 \\
    0 \: \: & \: \: 0 \: \: & \: \: 0 \: \: & \: \: 0 \: \: & \: \: \bar{T}^{\alpha}_{2,2} \: \: & \: \: 0
    \: \: & \: \: 0 \: \: & \: \: 0 \: \: & \: \: 0 \\
    0 \: \: & \: \: 0 \: \: & \: \: 0 \: \: & \: \: 0 \: \: & \: \: 0 \: \: & \: \: \bar{T}^{\alpha}_{2,3}
    \: \: & \: \: 0 \: \: & \: \: 0 \: \: & \: \: 0
  \end{matrix}
  \right)
\end{equation}
The index $\alpha$ runs over the generators of $U(N)$. In the
notation used above all the $T^{\alpha}_{i,j}$ for any $i,j$ are
in fact the same when considered as generators of $U(N)$, but the
notation is meant as a reminder that they act only as generators
of the $U(N)$ group at the lattice site denoted by $i,j$.

Diagonal elements correspond to adjoint fields and the off-diagonal elements to bi-fundamentals. The row and column indices
of the matrix can be derived by flattening the $i,j$ coordinates
via $i,j \rightarrow \left[ \left( i-1 \right) \times q \right] +
j$. In this notation we identify $i=0$ with $i=p$ and $j=0$ with
$j=q$. When writing things in matrix form, the fundamental index
indicates the row and the anti-fundamental index the column, for
both fields and their conjugates. When multiplying matrices, it is
important to note that
\begin{equation}
  \left( N_{i,j} \: , \: \bar{N}_{k,l} \right) \otimes
  \left( N_{k,l} \: , \: \bar{N}_{m,n} \right) \: \sim \:
  \left( N_{i,j} \: , \: \bar{N}_{m,n} \right) \ .
\end{equation}

Making use of these generators, we can now   explicitly construct
the one-loop correction to the dilatation operator. Starting with
\eqref{n4-dilatation} and evaluating the trace after the
orbifolding gives rise to a sum of terms, each situated at a
lattice site, and a trace over $N \times N$ matrices associated to
the gauge group sitting at the sites. This is a $3 \times 3$
triangular lattice, of the form discussed in section
\ref{the-lattice}. In general, there are many terms which appear
in this sum. These terms are more easily described in terms of
``interaction'' plaquettes on the lattice, as described in section
\ref{transfer-matrix}. We are primarily interested in terms which
give non-vanishing contributions when acting on purely holomorphic
operators. These arise from the first term in
\eqref{n4-dilatation}, with the second term giving similar
non-vanishing terms when acting on purely anti-holomorphic
operators. The other terms in $\mathfrak{D}_1$ have one
holomorphic and one anti-holomorphic derivative, and so vanish
when acting on operators with only holomorphic or only
anti-holomorphic fields. We summarize the terms of interest for
holomorphic operators \be \label{dilatation-example}
\begin{aligned}
  \mathfrak{D}_{1} =
  2 \: \sum_{i,j}
  tr
  \Big(
  \: \: & B_{i+1,j+1} A_{i,j} \Delta^A_{i,j} \Delta^B_{i+1,j+1}
\\
  \: - \: & A_{ij} B_{i,j+1} \Delta^A_{i-1,j-1} \Delta^B_{i,j}
\\
  \: - \: & B_{i+1,j+1} A_{i,j} \Delta^B_{i+1,j+2} \Delta^A_{i+1,j+1}
\\
  \: + \: & A_{i,j} B_{i,j+1} \Delta^B_{i,j+1} \Delta^A_{i,j}
\\
  \: + \: & C_{i-1,j} A_{i,j} \Delta^A_{i,j} \Delta^C_{i-1,j}
\\
  \: - \: & A_{i,j} C_{i,j+1} \Delta^A_{i+1,j} \Delta^C_{i,j}
\\
  \: - \: & C_{i-1,j} A_{i,j} \Delta^C_{i-1,j+1} \Delta^A_{i-1,j}
\\
  \: + \: & A_{i,j} C_{i,j+1} \Delta^C_{i,j+1} \Delta^A_{i,j}
\\
  \: + \: & C_{i-1,j} B_{i,j} \Delta^B_{i,j} \Delta^C_{i-1,j}
\\
  \: - \: & B_{i,j} C_{i-1,j-1} \Delta^B_{i+1,j} \Delta^C_{i,j}
\\
  \: - \: & C_{i-1,j} B_{i,j} \Delta^C_{i-2,j-1} \Delta^B_{i-1,j}
\\
  \: + \: & B_{i,j} C_{i-1,j-1} \Delta^C_{i-1,j-1} \Delta^B_{i,j}
  \Big) \ .
\end{aligned}
\ee%
The sum above is over all lattice sites, and we impose periodic
boundary conditions as before. Each field appearing above takes
values in the lie algebra of $U(N)$, and $tr$ denotes a trace over
the $U(N)$ indices. For example, $A_{i,j}$ is an $U(N)$ lie
algebra element sitting on the $A_{i,j}$ link, and under a gauge
transformation the field $A_{i,j}$ transforms as $A_{i,j}
\rightarrow U_{i,j} A_{i,j} U^{\dagger}_{i,j+1}$, with the
subscripts on the $U$'s denoting the gauge group whose
transformations they generate, as appropriate for bi-fundamental
fields. Similar rules apply to the other fields (see
\eqref{transforms} for their transformation properties). Using
these rules it is easy to see that each term in
\eqref{dilatation-example} is gauge invariant.

Examples of terms which only give non-zero
contributions when acting on operators with both holomorphic and
anti-holomorphic fields are
$tr(A_{i,j} \bar{\Delta}^B_{i,j+1} \bar{A}_{i-1,j-1} \Delta^B_{i,j})$
and
$tr(A_{i,j}
\Delta^B_{i+1,j+2} \bar{A}_{i+1,j+1} \bar{\Delta}^B_{i+1,j+1})$.
Note that for such operators it is not necessary for the solid
(likewise dashed) lines to touch each other. They generally lead
to $1/N$ non-planar interactions. Other examples are $tr(A_{i,j}
\bar{B}_{i+1,j+2} \Delta^A_{i+1,j+1} \bar{\Delta}^B_{i+1,j+1})$
which has a similar structure to the open plaquettes already
discussed, aside from the appearance of mixed fields, and new
``flat'' plaquettes of the form $tr(A_{i,j} \bar{A}_{i,j}
\Delta^A_{i,j-1} \bar{\Delta}^A_{i,j-1})$ and $tr(A_{i,j}
\bar{A}_{i,j} \bar{\Delta}^A_{i,j} \Delta^A_{i,j})$. These terms
are depicted in Figure \ref{non-hol}.

We now give two examples that demonstrates how an interaction plaquette acts on composite operators.
For this we need to know how the Wick contractions act. Following the definitions in the previous appendix, we
have
\be
  \left( \Delta^{I}_{kl} \right)^{ab}
  \left( \Phi^{J}_{mn} \right)^{cd} \: = \:
  \delta^{IJ} \ \delta_{km} \ \delta_{ln} \ \delta^{ad} \ \delta^{bc}
\ee
with $I,J$ ranging over $A,B,C$, $(k,l),(m,n)$ the lattice coordinates, and $(a,b),(c,d)$ $N \times N$
matrix indices.
Consider now an operator of the form
\be
  O = tr : \left( \bar{C}_{1,1} \bar{B}_{2,2} C_{2,2} B_{3,2} \right) : \ ,
\ee
together with the interaction term
\be
  I = tr : \left( \Delta^{B}_{3,2} \Delta^{C}_{2,2} B_{2,2} C_{1,1} \right) : \ ,
\ee
which is one possible interaction term of the form appearing in figure \ref{open-plaquettes}.
The interaction term $I$ operates on $O$ as follows
\be
  I \ O = N \: tr : \left( B_{2,2} C_{1,1} \bar{C}_{1,1} \bar{B}_{2,2} \right) : \ .
\ee
The factor of $N$ shows that this contraction is at planar level.
Here the single factor of $N$ arises from the contraction of two derivatives in a single trace from the interaction
with two fields in another single trace.

A planar example involving one of the plaquettes in Figure
\ref{non-hol} is given by%
\be
  I = tr : \left( A_{1,2} \bar{A}_{1,2} \Delta^{A}_{1,1} \bar{\Delta}^{A}_1,1 \right) :
\ee
and
\be
  O = tr : \left( A_{1,1} A_{1,2} \bar{A}_{1,2} \bar{A}_{1,1} \right) :
\ee
Then
\be
  I \ O = N \: tr : \left( A_{1,2} \bar{A}_{1,2} A_{1,2} \bar{A}_{1,2} \right) :
\ee%
For a final example that involves non-holomorphic operators as
well as non-planar interactions consider%
 \be
  O =
  tr : \left( A_{2,2} B_{2,3} \bar{A}_{1,1} \bar{C}_{2,2 }\right) :
  tr : \left( \bar{A}_{2,1} \bar{C}_{3,2} A_{3,2} B_{3,3} \right) : \ ,
\ee%
 which is a double trace operator representing two closed
strings on the lattice. Take the interaction term to be %
\be
  I = tr : \left( A_{2,2} \Delta^{A}_{2,2} \Delta^{B}_{3,3} B_{3,3} \right) : \ .
\ee
Using the above rules for contractions, we arrive at
\be
  I \ O = tr : \left( A_{2,2} B_{2,3} \bar{A}_{1,1} C_{2,2} \bar{A}_{2,1}
  \bar{C}_{3,2} A_{3,2} B_{3,3} \right) : \ .
\ee%
 This operator is suppressed by one power of $N$ relative to
the previous one, showing that it is non-planar, and involves the
joining of two strings into one. Here the non-planar dependence on
$N$ is due to the fact that the two derivatives inside the single
trace interaction operator  acts on two fields in different
traces, and it costs one factor of $1/N$ to join these traces.
This how strings join and split.

Finally, we would like to comment on the general case of an
asymmetric orbifold, where $p \ne q$. In this case the boundary
conditions break the $\mathbb{Z}_3$ rotational symmetry of the
lattice theory. The dilatation operator when expanded on the
lattice then also contains terms which are no longer invariant
under the $\mathbb{Z}_3$ symmetry relabeling the fields.



\begin{thebibliography}{99}



\bibitem{AdS/CFT}
  J.~M.~Maldacena,
  ``The large N limit of superconformal field theories and supergravity,''
  Adv.\ Theor.\ Math.\ Phys.\  {\bf 2}, 231 (1998)
  [Int.\ J.\ Theor.\ Phys.\  {\bf 38}, 1113 (1999)]
  [arXiv:hep-th/9711200].

  S.~S.~Gubser, I.~R.~Klebanov and A.~M.~Polyakov,
  ``Gauge theory correlators from non-critical string theory,''
  Phys.\ Lett.\ B {\bf 428}, 105 (1998)
  [arXiv:hep-th/9802109].

  E.~Witten,
  ``Anti-de Sitter space and holography,''
  Adv.\ Theor.\ Math.\ Phys.\  {\bf 2}, 253 (1998)
  [arXiv:hep-th/9802150].

\bibitem{Aharony:1999ti}
  O.~Aharony, S.~S.~Gubser, J.~M.~Maldacena, H.~Ooguri and Y.~Oz,
  ``Large N field theories, string theory and gravity,''
  Phys.\ Rept.\  {\bf 323}, 183 (2000)
  [arXiv:hep-th/9905111].



\bibitem{BMN}
  D.~Berenstein, J.~M.~Maldacena and H.~Nastase,
  ``Strings in flat space and pp waves from N = 4 super Yang Mills,''
  JHEP {\bf 0204}, 013 (2002)
  [arXiv:hep-th/0202021].

%

\bibitem{Plefka}
  J.~C.~Plefka,
  ``Lectures on the plane-wave string / gauge theory duality,''
  Fortsch.\ Phys.\  {\bf 52}, 264 (2004)
  [arXiv:hep-th/0307101].

\bibitem{BMNreview}
  D.~Sadri and M.~M.~Sheikh-Jabbari,
  ``The plane-wave / super Yang-Mills duality,''
  Rev.\ Mod.\ Phys.\  {\bf 76}, 853 (2004)
  [arXiv:hep-th/0310119].


\bibitem{Beisert:2003tq}
  N.~Beisert, C.~Kristjansen and M.~Staudacher,
  ``The dilatation operator of N = 4 super Yang-Mills theory,''
  Nucl.\ Phys.\ B {\bf 664}, 131 (2003)
  [arXiv:hep-th/0303060].

\bibitem{Beisert:2003jj}
  N.~Beisert,
  ``The complete one-loop dilatation operator of N = 4 super Yang-Mills
  theory,''
  Nucl.\ Phys.\ B {\bf 676}, 3 (2004)
  [arXiv:hep-th/0307015].

\bibitem{Minahan-Zarembo}
J.~A.~Minahan and K.~Zarembo, ``The Bethe-ansatz for N = 4 super
Yang-Mills,'' JHEP {\bf 0303}, 013 (2003) [arXiv:hep-th/0212208].

\bibitem{Beisert-review}
N.~Beisert, ``The dilatation operator of N = 4 super Yang-Mills
theory and integrability,'' Phys.\ Rept.\  {\bf 405}, 1 (2005)
[arXiv:hep-th/0407277].




\bibitem{Beisert:2003yb}
N.~Beisert and M.~Staudacher, ``The N = 4 SYM integrable super
spin chain,'' Nucl.\ Phys.\ B {\bf 670}, 439 (2003)
[arXiv:hep-th/0307042].


\bibitem{Kruczenski:2003gt}
S.~Frolov and A.~A.~Tseytlin,
``Multi-spin string solutions in AdS(5) x S**5,''
  Nucl.\ Phys.\ B {\bf 668} (2003) 77
  [arXiv:hep-th/0304255].

  G.~Arutyunov, S.~Frolov, J.~Russo and A.~A.~Tseytlin,
  ``Spinning strings in AdS(5) x S**5 and integrable systems,''
  Nucl.\ Phys.\ B {\bf 671} (2003) 3
  [arXiv:hep-th/0307191].

M.~Kruczenski, ``Spin chains and string theory,'' Phys.\ Rev.\
Lett.\  {\bf 93}, 161602 (2004) [arXiv:hep-th/0311203].

\bibitem{Kruczenski:2004wg}
M.~Kruczenski, ``Spiky strings and single trace operators in gauge
theories,'' arXiv:hep-th/0410226.


\bibitem{Bena:2003wd}
  I.~Bena, J.~Polchinski and R.~Roiban,
  ``Hidden symmetries of the AdS(5) x S**5 superstring,''
  Phys.\ Rev.\ D {\bf 69}, 046002 (2004)
  [arXiv:hep-th/0305116].


\bibitem{Beisert-higher-loops}
N.~Beisert, ``Higher loops, integrability and the near BMN
limit,'' JHEP {\bf 0309}, 062 (2003) [arXiv:hep-th/0308074];
N.~Beisert, ``The su(2$|$3) dynamic spin chain,'' Nucl.\ Phys.\ B
{\bf 682}, 487 (2004) [arXiv:hep-th/0310252].

\bibitem{Long-range}
N.~Beisert, V.~Dippel and M.~Staudacher, ``A novel long range spin
chain and planar N = 4 super Yang-Mills,'' JHEP {\bf 0407}, 075
(2004) [arXiv:hep-th/0405001].

 N.~Beisert and M.~Staudacher,
``Long-range PSU(2,2$|$4) Bethe ansaetze for gauge theory and
strings,''
  arXiv:hep-th/0504190.

\bibitem{Bethe}
H. Bethe, Z. Phys. {\bf 71}, 205 (1931).



\bibitem{higher-loop-failing}
J.~A.~Minahan, ``Higher loops beyond the SU(2) sector,'' JHEP {\bf
0410}, 053 (2004) [arXiv:hep-th/0405243].

C.~G.~.~Callan, T.~McLoughlin and I.~Swanson, ``Higher impurity
AdS/CFT correspondence in the near-BMN limit,'' Nucl.\ Phys.\ B
{\bf 700}, 271 (2004) [arXiv:hep-th/0405153].

T.~McLoughlin and I.~Swanson, ``N-impurity superstring spectra
near the pp-wave limit,'' Nucl.\ Phys.\ B {\bf 702}, 86 (2004)
[arXiv:hep-th/0407240].


\bibitem{QBethe}
 G.~Arutyunov, S.~Frolov and M.~Staudacher,
  ``Bethe ansatz for quantum strings,''
  JHEP {\bf 0410} (2004) 016
  [arXiv:hep-th/0406256].

 N.~Beisert and A.~A.~Tseytlin,
  ``On quantum corrections to spinning strings and Bethe equations,''
  arXiv:hep-th/0509084.

 J.~A.~Minahan, A.~Tirziu and A.~A.~Tseytlin,
  ``1/J corrections to semiclassical AdS/CFT states from quantum
 Landau-Lifshitz model,''
  arXiv:hep-th/0509071;
  ``1/J**2 corrections to BMN energies from the quantum long range
  Landau-Lifshitz model,''
  arXiv:hep-th/0510080.

  S.~Schafer-Nameki, M.~Zamaklar and K.~Zarembo,
  ``Quantum corrections to spinning strings in $AdS_5\times S^5$ and Bethe ansatz:
  A comparative study,''
  JHEP {\bf 0509} (2005) 051
  [arXiv:hep-th/0507189].

 S.~Schafer-Nameki and M.~Zamaklar,
  ``Stringy sums and corrections to the quantum string Bethe ansatz,''
  arXiv:hep-th/0509096.





\bibitem{selfdualYM}
L. Dolan, {\sl A New Symmetry Group of Real Self-dual Yang-Mills},
Phys. Lett. {\bf 113B} (1982) 273.\\
L.L.\ Chau, M.L.\ Ge and Y.S.\ Wu, {\sl Kac-Moody Algebra in the
Self-dual Yang-Mills Equation}, Phys. Rev. {\bf D25} (1982)
1086;\\ L.L.\ Chau and Y.S.\ Wu, {\sl More about Hidden Symmetry
Algebra for the Self-Dual Yang-Mills System} Phys. Rev.
{\bf D26} (1982) 3581; \\
L.L.\ Chau, M.L.\ Ge, A. Sinha and Y.S.\ Wu, {\sl Hidden Symmetry
Algebra for Self-Dual Yang-Mills Equations} Phys. Lett. {\bf 121B}
(1983) 391.\\
H.C.\ Tze and Y.S.\ Wu, {\sl Infinite Number of Local Conservation
Laws for the $SU(2)$ Self-dual Yang-Mills Systems}, Nucl. Phys.
{\bf B204} (1982) 118.\\
K. Ueno and Y. Nakamura, {\sl Transformation Theory for
(Anti)Self-dual Equations and the Riemann-Hilbert Problem}, Phys.
Lett. {\bf 109B} (1982) 273. \\ J. Avan, H.J. de Vega and J.M.
Maillet. {\sl Conformally Covariant Linear System for the
Four-Dimensional Self-Dual Yang-Mills Theory}, Phys. Lett. {\bf
171B} (1986) 255. \\ W.A.\ Bardeen, {\sl Selfdual Yang-Mills
Theory, Integrability and Multiparton Ampilitudes}, Prog. Theor.
Phys. Suppl. {bf 123} (1996) 1. \\ A.D. Popov and C.R.
Preitschopf, {\sl Conformal Symmetries of the Self-Dual Yang-Mills
Equations}, Phys.\ Lett. {\bf B374} (1996) 71; A.D. Popov, {\sl
Self-Dual Yang-Mills: Symmetries and Moduli Space}, Rev.\ Math.\
Phys. {\bf 11} (1999) 1091.





\bibitem{QCDspinchain}
L.N. Lipatov, {\sl High-energy Asymptotics of Multicolor QCD and
Exactly Solvable Lattice Models}, JETP Lett. {\bf 59} (1994)
596.\\
L.D.\ Faddev and G.P.\ Korchemsky, {\sl High-energy QCD as a
Completely Integrable Model}, Phys. Lett. {\bf B342} (1995) 311.\\
A.V. Belitsky, A.S. Gorsky and G.P. Korchemsky, {\sl Gauge /
String Duality For QCD Conformal Operators}, Nucl. Phys. {\bf
B667} (2003) 3. \\ V.M. Braun, S.E. Derkachov and A.N. Manashov,
{\sl Integrability Of Three Particle Evolution Equations In QCD},
Phys. Rev. Lett. {\bf 81} (1998) 2020; V.M. Braun, S.E. Derkachov,
G.P. Korchemsky and A.N. Manashov, {\sl Baryon Distribution
Amplitudes In QCD}, Nucl. Phys. {\bf B553} (1999) 355; A.V.
Belitsky, {\sl Renormalization Of Twist - Three Operators And
Integrable Lattice Models}, Nucl. Phys. {\bf B574} (2000) 407;
S.E. Derkachov, G.P. Korchemsky and A.N. Manashov, {\sl Evolution
Equations For Quark Gluon Distributions In Multicolor QCD And Open
Spin Chain}, Nucl. Phys. {\bf B566} (2000) 203.

\bibitem{QCDspinchain2}
  N.~Beisert, G.~Ferretti, R.~Heise and K.~Zarembo,
``One-loop QCD spin chain and its spectrum,''
  Nucl.\ Phys.\ B {\bf 717}, 137 (2005)
  [arXiv:hep-th/0412029].





\bibitem{KL03} A.V. Kotikov, L.N. Lipatov and V.N. Velizhanin,
{\sl Anomalous Dimensions of Wilson Operators in $\mathcal{N}=4$
SYM theory}, Phys. Lett. {\bf B557} (2003) 114.




\bibitem{Douglas:1996sw}
M.~R.~Douglas and G.~W.~Moore, ``D-branes, Quivers, and ALE
Instantons,'' arXiv:hep-th/9603167.

\bibitem{Eva-Shamit}
S.~Kachru and E.~Silverstein, ``4d conformal theories and strings
on orbifolds,'' Phys.\ Rev.\ Lett.\  {\bf 80}, 4855 (1998)
[arXiv:hep-th/9802183].

\bibitem{vafa-et.al.}
A.~E.~Lawrence, N.~Nekrasov and C.~Vafa, ``On conformal field
theories in four dimensions,'' Nucl.\ Phys.\ B {\bf 533}, 199
(1998) [arXiv:hep-th/9803015].




  Z.~Kakushadze,
  ``Gauge theories from orientifolds and large N limit,''
  Nucl.\ Phys.\ B {\bf 529}, 157 (1998)
  [arXiv:hep-th/9803214].

M.~Bershadsky and A.~Johansen, ``Large N limit of orbifold field
theories,'' Nucl.\ Phys.\ B {\bf 536}, 141 (1998)
[arXiv:hep-th/9803249].

\bibitem{general-quiver}
  S.~Benvenuti and M.~Kruczenski,
 ``From Sasaki-Einstein spaces to quivers via BPS geodesics: L(p,q$|$r),''
  arXiv:hep-th/0505206;
  ``Semiclassical strings in Sasaki-Einstein manifolds and long operators in $N
  = 1$ gauge theories,''
  arXiv:hep-th/0505046.





\bibitem{Wang:2003cu}
  X.~J.~Wang and Y.~S.~Wu,
  ``Integrable spin chain and operator mixing in N = 1,2 supersymmetric
  theories,''
  Nucl.\ Phys.\ B {\bf 683}, 363 (2004)
  [arXiv:hep-th/0311073].

\bibitem{Chen:2004mu}
  B.~Chen, X.~J.~Wang and Y.~S.~Wu,
  ``Integrable open spin chain in super Yang-Mills and the plane-wave / SYM
  duality,''
  JHEP {\bf 0402}, 029 (2004)
  [arXiv:hep-th/0401016].


\bibitem{Semenoff}
G.~De Risi, G.~Grignani, M.~Orselli and G.~W.~Semenoff, ``DLCQ
string spectrum from N = 2 SYM theory,'' arXiv:hep-th/0409315.

\bibitem{Compactified-ppwaves}
 M.~Bertolini, J.~de Boer, T.~Harmark, E.~Imeroni and N.~A.~Obers,
``Gauge theory description of compactified pp-waves,''
  JHEP {\bf 0301}, 016 (2003)
  [arXiv:hep-th/0209201].


\bibitem{Hanany:1997tb}
A.~Hanany and A.~Zaffaroni, ``On the realization of chiral
four-dimensional gauge theories using branes,'' JHEP {\bf 9805},
001 (1998) [arXiv:hep-th/9801134].

\bibitem{Orbifolds-Penrose-limit}
 N.~Itzhaki, I.~R.~Klebanov and S.~Mukhi,
``PP wave limit and enhanced supersymmetry in gauge theories,''
  JHEP {\bf 0203}, 048 (2002)
  [arXiv:hep-th/0202153].


  M.~Alishahiha and M.~M.~Sheikh-Jabbari,
  ``The pp-wave limits of orbifolded AdS(5) x S(5),''
  Phys.\ Lett.\ B {\bf 535}, 328 (2002)
  [arXiv:hep-th/0203018]

 N.~w.~Kim, A.~Pankiewicz, S.~J.~Rey and S.~Theisen,
 ``Superstring on pp-wave orbifold from large-N quiver gauge theory,''
  Eur.\ Phys.\ J.\ C {\bf 25}, 327 (2002)
  [arXiv:hep-th/0203080].




 K.~Oh and R.~Tatar,
``Orbifolds, Penrose limits and supersymmetry enhancement,''
  Phys.\ Rev.\ D {\bf 67}, 026001 (2003)
  [arXiv:hep-th/0205067].
 M.~Alishahiha, M.~M.~Sheikh-Jabbari and R.~Tatar,
``Spacetime quotients, Penrose limits and conformal symmetry restoration,''
  JHEP {\bf 0301}, 028 (2003)
  [arXiv:hep-th/0211285].


\bibitem{worldsheet-deconst.}
 S.~Mukhi, M.~Rangamani and E.~Verlinde,
 ``Strings from quivers, membranes from moose,''
  JHEP {\bf 0205}, 023 (2002)
  [arXiv:hep-th/0204147].

M.~Alishahiha, M.~M.~Sheikh-Jabbari,
``Strings in PP-waves and worldsheet deconstruction,''
  Phys.\ Lett.\ B {\bf 538}, 180 (2002)
  [arXiv:hep-th/0204174].





\bibitem{Kogut}
J.~B.~Kogut, ``An Introduction To Lattice Gauge Theory And Spin
Systems,'' Rev.\ Mod.\ Phys.\  {\bf 51}, 659 (1979).



\bibitem{fractional-brane}
  D.~E.~Diaconescu, M.~R.~Douglas and J.~Gomis,
  ``Fractional branes and wrapped branes,''
  JHEP {\bf 9802}, 013 (1998)
  [arXiv:hep-th/9712230].



%
\bibitem{Wess:1992cp}
  J.~Wess and J.~Bagger,
  ``Supersymmetry and supergravity,''
Princeton University Press; 2nd Rev. edition (March 3, 1992).



\bibitem{HW}
A.~Hanany and E.~Witten,
``Type IIB superstrings, BPS monopoles, and three-dimensional gauge
dynamics,''
Nucl.\ Phys.\ B {\bf 492}, 152 (1997)
[arXiv:hep-th/9611230].

\bibitem{Hanany-Uranga}
A.~Hanany and A.~M.~Uranga, ``Brane boxes and branes on
singularities,'' JHEP {\bf 9805}, 013 (1998)
[arXiv:hep-th/9805139].

\bibitem{KR81}P.R.\ Kulish and N.Yu.\ Reshetikin, {\sl Generalized
Heisenberg Ferromagnet and the Gross-Neveu model}, Sov. Phys. JETP
{\bf 53} (1981) 108.
\bibitem{Faddeev96}L.D.\ Faddeev, {\sl How Algebraic Bethe Ansatz
Works for Integrable Model}, hep-th/9605187.




\bibitem{Aharony:1997bh}
O.~Aharony, A.~Hanany and B.~Kol,
``Webs of (p,q) 5-branes, five dimensional field theories and grid
diagrams,''
JHEP {\bf 9801}, 002 (1998)
[arXiv:hep-th/9710116].




\bibitem{Erlich:1999rb}
J.~Erlich, A.~Hanany and A.~Naqvi,
``Marginal deformations from branes,''
JHEP {\bf 9903}, 008 (1999)
[arXiv:hep-th/9902118].

\bibitem{Feng:1999fw}
B.~Feng, A.~Hanany and Y.~H.~He,
``The Z(k) x D(k') brane box model,''
JHEP {\bf 9909}, 011 (1999)
[arXiv:hep-th/9906031].

\bibitem{Feng:1999zv}
B.~Feng, A.~Hanany and Y.~H.~He,
``Z-D brane box models and non-chiral dihedral quivers,''
arXiv:hep-th/9909125.



\bibitem{Hellerman:2002qa}
S.~Hellerman, ``Lattice gauge theories have gravitational duals,''
arXiv:hep-th/0207226.

\bibitem{Mithat}
 D.~B.~Kaplan, E.~Katz and M.~Unsal,
 ``Supersymmetry on a spatial lattice,''
  JHEP {\bf 0305}, 037 (2003)
  [arXiv:hep-lat/0206019].

  D.~B.~Kaplan and M.~Unsal,
  ``A Euclidean lattice construction of supersymmetric Yang-Mills theories
  with sixteen supercharges,''
  JHEP {\bf 0509}, 042 (2005)
  [arXiv:hep-lat/0503039].

 M.~Unsal,
  ``Regularization of non-commutative SYM by orbifolds with discrete torsion
  and SL(2,Z) duality,''
  arXiv:hep-th/0409106.







\bibitem{Arkani-Hamed:2001ie}
N.~Arkani-Hamed, A.~G.~Cohen, D.~B.~Kaplan, A.~Karch and
L.~Motl,``Deconstructing (2,0) and little string theories,'' JHEP
{\bf 0301}, 083 (2003) [arXiv:hep-th/0110146].

\bibitem{Halpern}
  M.~B.~Halpern and W.~Siegel,
  ``Electromagnetism As A Strong Interaction,''
  Phys.\ Rev.\ D {\bf 11}, 2967 (1975).






\bibitem{dimer}
  A.~Hanany and K.~D.~Kennaway,
 ``Dimer models and toric diagrams,''
  arXiv:hep-th/0503149.

  S.~Franco, A.~Hanany, K.~D.~Kennaway, D.~Vegh and B.~Wecht,
  ``Brane dimers and quiver gauge theories,''
  arXiv:hep-th/0504110.

S.~Franco, A.~Hanany, D.~Martelli, J.~Sparks, D.~Vegh and B.~Wecht,
``Gauge theories from toric geometry and brane tilings,''
  arXiv:hep-th/0505211.


 T.~Muto,
  ``A relation between moduli space of D-branes on orbifolds and Ising model,''
  arXiv:hep-th/0508248.







\end{thebibliography}
\end{document}